\documentclass[%
 reprint, 
superscriptaddress,
nofootinbib,
 amsmath,amssymb,
 aps, physrev,
 prl,
]{revtex4-2}

\usepackage{graphicx}
\usepackage{dcolumn}
\usepackage{bm}
\usepackage[dvipsnames]{xcolor}
\usepackage{hyperref}
\hypersetup{
	colorlinks=true,
	linkcolor=Maroon,
	citecolor=OliveGreen,
	urlcolor=MidnightBlue,
}
\usepackage{booktabs}
\usepackage{multirow}
\usepackage{acronym}
\usepackage{xspace}
\usepackage[capitalize]{cleveref}

\usepackage{orcidlink}
 \newcommand{\orcid}[1]{\href{https://orcid.org/#1}{\includegraphics[width=10pt]{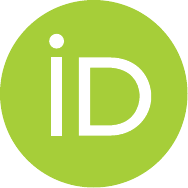}}}

\begin{document}


\title{The fault in our sirens: Hierarchical diagnosis of waveform systematics\\ in Hubble-Lema\^itre--constant measurements}

\newcommand{\eob}{\texttt{SEOBNRv5PHM}\xspace}
\newcommand{\xphm}{\texttt{IMRPhenomXPHM}\xspace}
\newcommand{\xoa}{\texttt{IMRPhenomXO4a}\xspace}
\newcommand{\seob}{\texttt{SEOBNR}\xspace}
\newcommand{\teob}{\texttt{TEOBResumS}\xspace}
\newcommand{\bilby}{\texttt{Bilby}\xspace}
\newcommand{\gwbench}{\texttt{GWBENCH}\xspace}
\newcommand{\mass}{\texttt{Power Law + Peak}\xspace}
\newcommand{\spin}{\texttt{DEFAULT}\xspace}
\newcommand{\dynesty}{\texttt{dynesty}\xspace}
\newcommand{\bilbypipe}{\texttt{Bilby-pipe}\xspace}
\newcommand{\hubble}{\ac{$H_0$}\xspace}
\newcommand{\lal}{\texttt{LALSuite}\xspace}
\newcommand{\mice}{\texttt{MICECAT}\xspace}
\newcommand{\HLV}{\text{O5}\xspace}
\newcommand{\As}{\text{A\#}\xspace}
\newcommand{\XG}{\text{XG}\xspace}
\newcommand{\glade}{\texttt{GLADE+}\xspace}

\newcommand{\error}[3]{#1^{+#2}_{-#3}}
\newcommand{\jg}[1]{\textcolor{Magenta}{[\sf{Jon: #1}]}}
\definecolor{cadmiumgreen}{rgb}{0.0, 0.42, 0.24}
\definecolor{cerulean}{rgb}{0.0, 0.48, 0.65}
\newcommand{\ad}[1]{\textcolor{cerulean}{\textbf{#1}}}
\newcommand{\ab}[1]{\textcolor{cyan}{\textbf{#1}}}

\newcommand{\aei}{\affiliation{Max Planck Institute for Gravitational Physics (Albert Einstein Institute), Am Mühlenberg 1, Potsdam 14476, Germany}}
\newcommand{\umd}{\affiliation{Department of Physics, University of Maryland, College Park, MD 20742, USA}}

\author{Arnab Dhani\,\orcid{0000-0001-9930-9101}}
\email{arnab.dhani@aei.mpg.de}
\aei

\author{Jonathan Gair\,\orcid{0000-0002-1671-3668}}%
\aei

\author{Alessandra Buonanno\,\orcid{0000-0002-5433-1409}}%
\aei
\umd

\begin{abstract}
	Cosmological inference using a population of binary black-hole mergers, combined with a galaxy catalog, presents an exciting 
	opportunity for precision cosmology with the possibility of resolving the Hubble tension. 
	However, the accuracy of these measurements heavily relies on the quality of the model used to infer the binary parameters, 
	including the model of the gravitational-wave signal.
	We use state-of-the-art waveform models to explore the impact of inaccurate modeling in measuring the Hubble-Lema\^itre constant 
	for the upcoming and future ground-based gravitational-wave observatories.
	We diagnose the presence of inaccuracies within a hierarchical population-analysis framework, without a priori 
	knowing the true value of the parameter, by assessing the consistency of the distribution of individual posteriors 
	in relation to their measurement errors. 
	Our findings indicate that even a small high-mass, spin-precessing subpopulation---comprising as little as 5\% of the 
	population generating the events observed by the LIGO-Virgo-KAGRA Collaboration so far---can 
	result in an unreliable measurement of the Hubble-Lema\^itre constant in the upcoming observing runs of these detectors, 
	with even more pronounced effects expected in future facilities on the ground.
\end{abstract}

\maketitle

\acrodef{BNS}{binary neutron star}
\acrodef{BBH}{binary black hole}
\acrodef{NSBH}{neutron-star--black-hole}
\acrodef{NS}{neutron star}
\acrodef{BH}{black hole}
\acrodef{GW}{gravitational-wave}
\acrodef{CBC}{compact-binary coalescence}
\acrodef{EM}{electromagnetic}
\acrodef{$H_0$}{\textrm{Hubble-Lema\^{i}tre parameter}}
\acrodef{XG}{next-generation}
\acrodef{CE}{Cosmic Explorer}
\acrodef{ET}{Einstein Telescope}
\acrodef{PISN}{pair-instability supernova}
\acrodef{PSD}{power spectral density}
\acrodef{SNR}{signal-to-noise ratio}
\acrodef{ADM}{Arnowitt-Deser-Misner}
\acrodef{LSA}{linear-signal approximation}
\acrodef{IMR}{inspiral-merger-ringdown}
\acrodef{LVK}{LIGO-Virgo-KAGRA}
\acrodef{EoS}{equation of state}
\acrodef{GR}{general relativity}
\acrodef{NR}{numerical relativity}
\acrodef{PN}{post-Newtonian}
\acrodef{PM}{post-Minkowskian}
\acrodef{SFR}{star formation rate}
\acrodef{EOB}{effective-one-body}
\acrodef{FIM}{Fisher information matrix}
\acrodef{GWTC}{Gravitational-Wave Transient Catalog}
\acrodef{CMB}{Cosmic Microwave Background}
\acrodef{HMP}{high-mass precessing}
\acrodef{MAP}{maximum a-posteriori}
\acrodef{HPD}{highest posterior density}
\acrodef{PPHD}{posterior predictive hyperdistribution}

\textit{Introduction---}
\Ac{GW} cosmology has emerged as a promising independent method for probing the cosmic expansion history of the Universe~\cite{LIGOScientific:2017adf,LIGOScientific:2021aug}. 
\Acp{GW} from \acp{CBC} serve as standard sirens that measure distance without relying on 
the traditional cosmic distance ladder.
When combined with redshift measurements, these distances can be used to estimate various cosmological parameters~\cite{1986Natur.323..310S}.
Several methods exist to obtain the redshift information of 
a \ac{GW} event.
The most straightforward approach is when an identifiable \ac{EM} counterpart localizes the host galaxy 
of the \ac{GW} event, as demonstrated for GW170817 \cite{LIGOScientific:2018gmd}. 
However, most \ac{GW} events do not have an associated \ac{EM} counterpart. 
In such cases, a statistical measurement of the \ac{$H_0$} can be obtained using a galaxy catalog of probable hosts, either by directly using the measured galaxy redshifts as a prior on the GW source redshifts~\cite{1986Natur.323..310S,DelPozzo:2011vcw,Borhanian:2020vyr}, or by cross-correlation the inferred statistical distribution of galaxies and that of GW sources~\cite{PhysRevD.103.043520,Mukherjee:2022afz}. 
Alternatively, redshift measurements can be obtained by 
examining the features in the mass, and indirectly the spin, distributions of the \ac{CBC} population~\cite{Chernoff:1993th,Taylor:2012db,Farr:2019twy,Ezquiaga:2022zkx,Tong:2025xvd}, 
or analyzing the stochastic \ac{GW} background~\cite{Cousins:2025bas}. 
Additionally, tidal measurements from a \ac{BNS} merger can provide the redshift to such \ac{GW} sources~\cite{Messenger:2011gi,Messenger:2013fya,Li:2013via,Dhani:2022ulg}.

With close to 100 observations of \acp{CBC} to date, the current estimate of \ac{$H_0$} using \acp{GW} is
$H_0 = 68^{+8}_{-6} \rm km s^{-1} Mpc^{-1}$ \cite{LIGOScientific:2021aug}. 
The ongoing observing run (O4) of the \ac{LVK} Collaboration~\cite{LIGOScientific:2014pky,VIRGO:2014yos,KAGRA:2020tym} 
has already detected $>$200 signal candidates, 
and this number is expected to grow to thousands as the detectors reach their design sensitivity by the end 
of the decade~\cite{Borhanian:2022czq}. 
Additionally, proposed future detectors, like the \ac{CE} \cite{Reitze:2019iox,Evans:2021gyd,Srivastava:2022slt} 
and the \ac{ET} \cite{Punturo:2010zz,Maggiore:2019uih,Branchesi:2023mws}, are anticipated to detect every stellar-origin 
\ac{BBH} merger in the Universe and every \ac{BNS} merger up to a redshift of 1, 
increasing the observed population of \acp{CBC} to hundreds of thousands~\cite{Borhanian:2022czq}. 
Since the measurement precision improves as the inverse square root of the number of observations,
$H_0$ measurements using \acp{GW} could achieve sub-percent precision \cite{Borhanian:2020vyr,Dhani:2022ulg,Muttoni:2023prw},
positioning \ac{GW} cosmology firmly in the realm of precision cosmology~\cite{Planck:2018vyg,Riess:2021jrx}. 

Given the expected precision of these measurements, it is essential that they are also accurate. 
Accurate measurements are particularly critical because \ac{GW} measurements aim to resolve 
the $\sim$$5\sigma$ discrepancy in 
$H_0$ estimated in the 
local-Universe 
and in the pre-recombination era~\cite{Planck:2018vyg,Riess:2021jrx}.

Accurate waveform models are essential for an unbiased inference of a \ac{GW} source's binary parameters.
Thanks to the advancements in \ac{GW} modeling techniques, current state-of-the-art waveform models 
can effectively analyze most signals found in the \ac{LVK} \acp{GWTC}~\cite{Owen:2023mid}.
However, as detector sensitivities continue to increase, the accuracy of these waveform models will be challenged, 
particularly for binaries with large masses, asymmetries, and spins~\cite{Dhani:2024jja,Kapil:2024zdn}. 
Consequently, any systematic errors in the waveform models would affect the inference of the properties of the \ac{BBH} population 
and likely impact the estimation of $H_0$~\cite{Purrer:2019jcp,Dhani:2024jja}.

While a direct \ac{EM} counterpart accompanying a \ac{GW} event would provide the most accurate measurement of $H_0$, 
such events are infrequent; to date, only one has been observed~\cite{LIGOScientific:2017vwq,LIGOScientific:2017adf}. 
Therefore, ``dark sirens'' consisting of loud \ac{BBH} mergers with no EM counterpart are expected to yield the most precise measurements, 
as there are likely to be few probable host galaxies within the localization volume~\cite{Borhanian:2020vyr,Muttoni:2023prw}. 
Previous studies of this method have focused on how assumptions regarding the population model and galaxy weights affect the measurement accuracies
\cite{LIGOScientific:2021aug,Hanselman:2024hqy}. 
However, none has systematically investigated the impact of inaccurate waveform models on the measurement of $H_0$ with this method. 

In this \textit{Letter}, we analyze the impact of inaccurate waveform models on the $H_0$ measurement 
using a population of simulated \ac{BBH} mergers and host galaxies observed 
in current and future ground-based \ac{GW} detectors. 
The current estimates from GWTC-3 indicate that most merging \acp{BBH} have relatively low masses, ranging from 10 to 30$M_{\odot}$, 
exhibit small asymmetries, and have small spins with minimal, if any, spin-precession~\cite{KAGRA:2021duu}. 
However, there are outliers, such as GW190521\_030229~\cite{LIGOScientific:2020iuh} 
and the candidate events GW190403\_051519 and GW200208\_222617~\cite{KAGRA:2021vkt},
that display larger masses, greater asymmetries, or higher spins, or some combination of these. 
While, the latter two events represent marginal detections with moderate astrophysical significance, 
the scientific potential of these events is substantial. 
Confident detections of similar events in the future will be invaluable and there are reasons to expect these to be observed.
Although the current set of \ac{GW} observations does not definitively demonstrate that \ac{BH} masses evolve with redshift~\cite{Rinaldi:2023bbd,Lalleman:2025xcs}, 
astrophysical considerations tend to support the idea that they should~\cite{Vink:2020nak,Weatherford:2021zdf,Zevin:2020gbd}, 
with larger masses, greater asymmetries, and higher spins expected at larger redshifts~\cite{Bavera:2022mef,Biscoveanu:2022qac,Ye:2024ypm}.
Moreover, strong lensing of \acp{GW} is expected to shift heavy systems at higher redshifts, nominally outside the detector horizon, into the detectable range. 
Unfortunately, these scientifically exciting systems 
are more challenging to model because \ac{NR} simulations in these regions of the parameter space are limited. 
Consequently, significant systematic biases can arise when analyzing these systems~\cite{Dhani:2024jja,Purrer:2019jcp}. 
For this reason, it is vital to take into account the effect of unobserved, yet anticipated, populations of \acp{BH}  
when evaluating the impact of inaccurate waveform models on \ac{GW} science and assessing the accuracy requirements for future waveform models. 
In this study, we embellish the population of \ac{BBH} mergers as observed by the \ac{LVK} Collaboration 
by including a small fraction of binaries with large masses, high asymmetries, and rapid spins.

\ac{$H_0$} is not a direct \ac{GW} observable, but, rather, a characteristic of the population of \ac{GW} sources. 
Therefore, a hierarchical Bayesian inference method is used to infer it from a distribution of sources. 
In an unbiased hierarchical inference, the individual \ac{$H_0$} posteriors should be distributed around the true value 
according to the noise properties. 
This is similar to how the posteriors on the parameters of individual \ac{GW} events fluctuate around the true value for different noise realizations.
However, when systematic biases are present --- such as an incorrect noise model or waveform model --- the consistency 
between the distribution of the individual posteriors and the noise properties may not hold. 
Indeed, the 
probability-probability (P-P) plot is widely used in \ac{GW} astronomy~\cite{Gair:2015nga} to asses this consistency for the parameters of individual sources. 
\textcite{Hanselman:2024hqy} developed a diagnostic tool to utilize this consistency condition for population parameters by going a layer deeper in the hierarchical analysis, 
treating the individual \ac{$H_0$} posteriors as coming from a population.
For Gaussian-distributed posteriors on the parameters of individual events whose values are  themselves drawn from a normally distributed population model, the joint constraint on the mean converges to the mean of the population, with an uncertainty that depends on both the individual event uncertainties and the variance in the population. Differences between the population variance and the individual event uncertainties are then indicative of the spread of parameter values in the population. 
Applying this approach to a parameter that should be common to all events is a powerful method for identifying systematic biases because it is constructed within the analysis framework, 
eliminating the need for comparisons that involve different models and assumptions. 
We employ this method to detect systematic biases that arise from inaccurate waveform models. We find that the method is able to detect the presence of systematic biases, even in situations in which the joint posterior converges to the true value (which would in any case not be known a priori).

\textit{Diagnosing population systematics---}
We briefly formulate the method for identifying systematic biases introduced in \textcite{Hanselman:2024hqy} in a simplified setting. 
While the simplifications allow us to obtain simple analytical results, the observations remain valid even in the general case.  
Let us assume we have $N$ noisy measurements of a parameter $\lambda$ obtained from data $\mathcal{D}$. 
This gives us a set of likelihoods represented as $\{p(\mathcal{D}_i|\lambda^i)\}_{i=1}^{N}$.
We will assume that each $\lambda$ is drawn from a population characterized by a Normal distribution,
$p(\lambda|\mu,\sigma) \sim \mathcal{N}(\mu,\sigma^2)$ where $\mu$ is the mean and $\sigma^2$ is the variance. 
In the hierarchical Bayesian inference framework, the likelihood function on $\{\mu,\sigma\}$ is given by
\begin{equation}
\label{eq:pop_hiermodel}
	p(\{\mathcal{D}_i\}|\mu,\sigma) = \prod_{i=1}^{N} \int d\lambda^i p(\mathcal{D}_i|\lambda^i) p(\lambda^i|\mu,\sigma).
\end{equation}
To understand the structure of the distribution mentioned above, it is useful to evaluate the maximum likelihood estimator. 
Assuming that the $\lambda$ measurements are normally distributed with mean $\hat{\lambda}^i$ and variance $\sigma^2_i$ 
(i.e., $p(\mathcal{D}_i|\lambda^i) \sim \mathcal{N}(\hat{\lambda}^i,\sigma_i^2)$), 
the maximum likelihood estimates for $\{\mu,\sigma\}$, denoted by $\{\hat{\mu},\hat{\sigma}\}$, 
is given by the solution of the following set of coupled equations~\cite{Isi:2022cii} 
\begin{equation}
	\begin{aligned}
		\hat{\mu} &= \frac{\sum_{i=1}^{N} w_i \hat{\lambda}^i}{\sum_{i=1}^{N} w_i}, \\
		\sum_{i=1}^{N} w_i &- \sum_{i=1}^{N} w_i^2 (\hat{\lambda}^i - \hat{\mu})^2 = 0,
	\end{aligned}
\end{equation}
where, $w_i = 1/(\sigma^2_i+\hat{\sigma}^2)$.
There are no closed-form solutions to this set of equations. 
Therefore, under the simplifying assumption that the errors in all the individual observations of $\lambda$ are the same, 
$\sigma_i=\sigma_{\lambda}$, we find
\begin{equation}
	\begin{aligned}
		\label{eq:popvar_indvar}
		\hat{\mu} &= \frac{1}{N} \sum_{i=1}^{N} \hat{\lambda}^i \\ 
		\hat{\sigma}^2 &= \frac{1}{N} \sum_{i=1}^{N} (\hat{\lambda}^i - \hat{\mu})^2 - \sigma_{\lambda}^2.
	\end{aligned}
\end{equation}
In particular, \cref{eq:popvar_indvar} indicates that the maximum likelihood estimate of the population mean 
is the average of the individual observation means. 
It further states that the population variance is the 
difference the scatter of the individual posteriors ($\sum_{i=1}^{N} (\hat{\lambda} - \hat{\mu})^2 / N$) and 
the variance of the measurements ($\sigma_{\lambda}^2$).

\begin{figure}[h!]
	\includegraphics[width=\columnwidth]{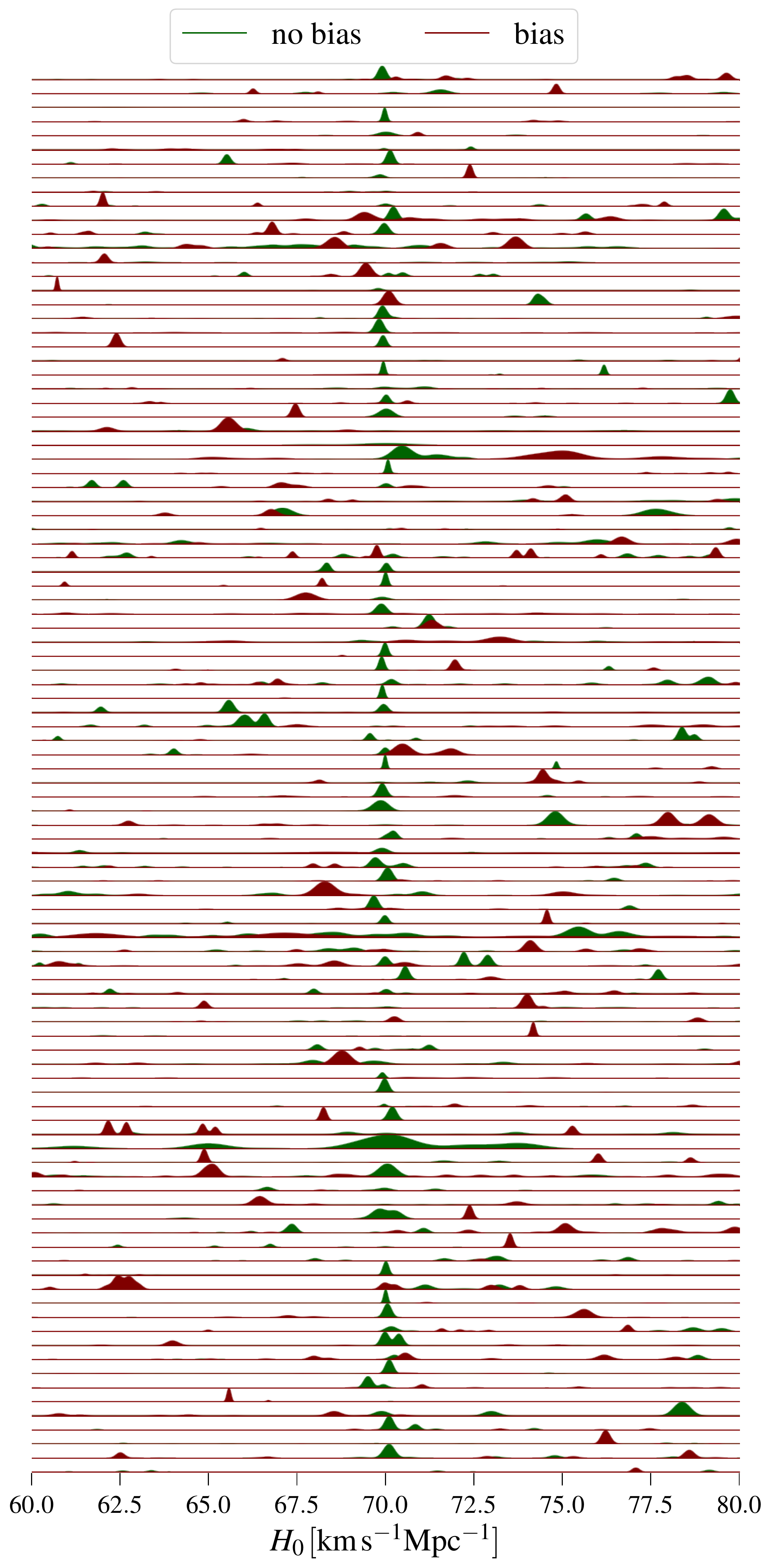}
	\caption{The individual $H_0$ posteriors in the absence (green) and presence (red) of systematic bias of a random selection 
		of 100 events with $\rm SNR>600$ for the high-mass spin-precessing population. 
		It also illustrates the practical difficulty in combining the posteriors in the presence of systematic biases. 
		The product of two narrow PDFs can be zero at machine precision using floating point arithmetic.
		}
	\label{fig:individual_H0}
\end{figure}

Suppose the true population follows a $\delta$-function, i.e., $\lambda$ is common across all observations, then 
$\sigma \rightarrow 0$ by construction. 
This is the case for \ac{$H_0$} since it is a constant of nature.
In this case, \cref{eq:popvar_indvar} asserts that $\hat{\sigma}$, on average, should be 0, and 
the scatter of the individual observations follows the measurement 
uncertainties. 
On the other hand, if $\sigma$ is measured to be non-zero, this indicates the presence of 
systematic bias in the computation of \ac{$H_0$}. 
In a hierarchical Bayesian analysis, if the likelihood does not follow the true data-generating process --- 
whether due to inaccuracies in the model, assumptions about the population, or characteristics of the noise --- 
the posterior parameters will not only converge to a biased value as the number of observations increase, 
but the scatter of the individual observations will also be inconsistent with the assumed noise properties.
While the first of these cannot be used as a diagnostic, since the true values of the parameters of interest are not known a priori, 
the scatter of the individual posteriors can serve as a diagnostic of model inconsistency. 
This is the key role of $\sigma$ in \cref{eq:pop_hiermodel}.
However, it is important to note that the source of the systematic error cannot be isolated. 
For instance, in this study, we examine systematic biases due to inaccurate waveform models, while 
\textcite{Hanselman:2024hqy} explored systematic biases due to inaccurate galaxy weights. 
We illustrate this in Fig.~\ref{fig:individual_H0} where it is clear that in the absence of systematic biases due to 
inaccurate waveform models, the $H_0$ posteriors cluster around the injected value of $70 \rm \, km \, s^{-1} \, Mpc^{-1}$
with a spread typical of the widths of the posteriors. 
In contrast, the posteriors affected by systematic biases are scattered randomly across the domain.
\cref{fig:individual_H0} also illustrates the practical difficulty in combining the posteriors in the presence of systematic biases 
where the product of two narrow and separated PDFs can be zero at machine precision using floating point arithmetic. 
However, this issue is mitigated by treating the \ac{$H_0$} values as arising from a population as is done in this paper
because we are no longer enforcing consistency in the means of the posteriors. 
We observe that, even when the models used for data generation and for the analysis are consistent, there are rare instances in which the primary mode of the posterior is significantly separated from the true value. However, when this happens there is always 
a secondary mode that aligns with the true value. 
This is not unexpected, as galaxies or galaxy clusters are likely to be found within the localization volume, even if they did not host the \ac{GW} event.

\begin{figure}
	\includegraphics[width=\columnwidth]{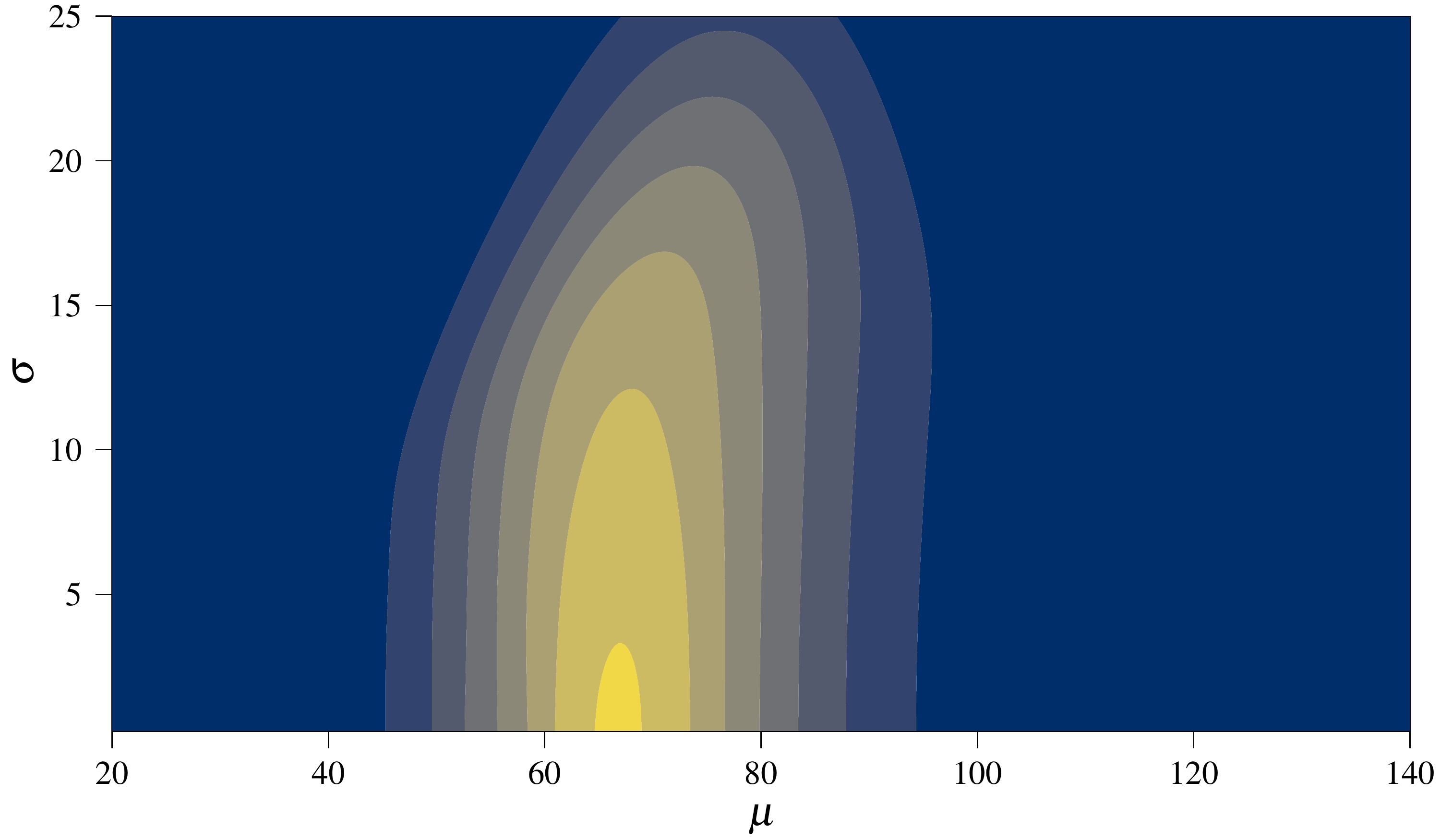}
	\caption{The posterior distribution of the hyperparameters $\{\mu, \sigma\}$ for the real events 
	used in the GWTC-3 cosmology paper.}
	\label{fig:gwtc-3_mu_sigma}
\end{figure}

We apply the hierarchical framework on the 46 real events used in~\textcite{LIGOScientific:2021aug}. 
We use the publicly available $H_0$ posteriors of each event for our analysis.
We choose the specific case corresponding to Fig. 7 of \textcite{LIGOScientific:2021aug}.
The 2D posterior on $\{\mu, \sigma\}$ is shown in Fig.~\ref{fig:gwtc-3_mu_sigma}. 
We find that $\sigma$ is consistent with zero and, therefore, 
the $H_0$ inference is not obviously affected by systematic biases.
We confirm this observation for other population and galaxy catalog choices that were considered in \textcite{LIGOScientific:2021aug}
and for which individual event posteriors are publicly available.
While the contours move around on the $\mu$ axis signifying that the particular choices affect the measurement, 
as seen in Fig. 11 of \textcite{LIGOScientific:2021aug},
we find that $\sigma$ is consistent with zero in all the cases indicating that the distribution of the individual 
event posteriors is consistent with the assumptions of the models. 
We emphasize that, for these real events, the $\sigma$ parameter is a test for any kind of systematics in the 
$H_0$ hierarachical likelihood and is not restricted to waveform systematics alone.

\textit{Systematic biases in upcoming observations---}
The current \ac{GW} observations do not exhibit any systematic biases in the inference of $H_0$. 
Therefore, we investigate the potential impact of systematic biases in both planned and future ground-based \ac{GW} observatories.
We simulate \ac{GW} observations across three detector networks.
These are (i) the \HLV network, which consists of the two LIGO detectors and the Virgo detector, all operating at their design sensitivity; 
(ii) the \As network where the LIGO detectors operating at an upgraded A\# sensitivity, 
while the Virgo detector continues to operate at design sensitivity; and 
(iii) the \XG network made up of two \ac{CE} detectors with arm lengths of 40 km and 20 km, 
along with a triangular \ac{ET} with an arm length of 10 km.
These networks represent highly probable observing scenarios for the future.
The \HLV network is expected to operate until the end of this decade, 
while the \As network is anticipated to begin functioning in the early part of the next decade, 
and the \XG network is expected to be operational in the latter half of that decade.

We simulate a population of \ac{BBH} mergers whose mass and spin distributions follow the median estimates from GWTC-3~\cite{KAGRA:2021vkt,KAGRA:2021duu}. 
We embellish this population with varying fractions of a high-mass spin-precessing subpopulation 
(see \cref{sec:HMP_population} for details). 
We note that the fractions reported in this paper correspond to the full population and not the detected population. 
The corresponding fractions in the detected population depend on the \ac{SNR} threshold and are impacted by the 
selection bias. 

We assume that the redshift distribution of the \ac{GW} sources follows the distribution of galaxies. 
To assign host galaxies and their corresponding redshifts and sky locations, 
we randomly select from the \mice mock galaxy catalog, 
which is complete and extends to a 
redshift of 1.4. 
We simulate a population of 100,000 binaries which correspond to a 4-yr observational period based on the 
estimates of the current merger rates~\cite{Borhanian:2022czq,KAGRA:2021duu}.
We apply a network \ac{SNR} threshold of 70, 100, 600 for the three detector networks \HLV, \As, and \XG, respectively.
We chose these high \ac{SNR} cutoffs to ensure that the GW detection horizon remains within that of the galaxy catalog.
This is crucial to prevent the edge of the galaxy catalog --- which contains information about \ac{$H_0$} --— 
from skewing our estimates, as it could result in artificially lower error estimates for \ac{$H_0$}.
Additionally, the number of potential host galaxies increases rapidly with the localization volume, 
which makes \ac{$H_0$} inference computationally demanding. 
Therefore, our decision to analyze events with higher \ac{SNR} values helps to keep the computations manageable. 
While accounting for \ac{EM} selection effects could allow for smaller \ac{SNR} thresholds, we opted for this approach for simplicity. 
Moreover, events with higher \ac{SNR} are the most informative, making it sufficient to analyze only this case.

\begin{table}
	\centering
	\renewcommand{\arraystretch}{1.3}
	\caption{The \ac{MAP} value 
		and the 90\% \ac{HPD} interval for the hyperparameters $\{\mu, \sigma\}$ 
		in the absence and presence of systematic biases in the different detector networks and the various
		fractions of the embellishments to the LVK population considered here. 
		The values quoted are median values over 100 realizations for each case.}
	\hspace*{-4mm}
	\begin{tabular}{@{}lrrcrrcrr@{}}
		\toprule
		\multirow{2}{*}{frac.} & \multicolumn{2}{c}{\HLV} && \multicolumn{2}{c}{\As} && \multicolumn{2}{c}{\XG} \\
		\cmidrule{2-3} \cmidrule{5-6} \cmidrule{8-9}
		& no-bias & bias && no-bias & bias && no-bias & bias \\
		\midrule
		$p(\mu)$ \\
		0\%  & $\error{69.5}{1.59}{1.33}$ & $\error{69.2}{1.52}{1.47}$ && $\error{70.2}{0.57}{0.45}$ & $\error{70.4}{0.44}{0.47}$ && $\error{70.0}{0.01}{0.02}$ & $\error{69.8}{0.09}{0.09}$ \\
		2\%  & $\error{69.6}{1.24}{1.16}$ & $\error{69.0}{1.51}{1.85}$ && $\error{70.1}{0.50}{0.39}$ & $\error{70.4}{0.56}{0.68}$ && $\error{70.0}{0.01}{0.02}$ & $\error{69.7}{0.11}{0.11}$ \\
		5\%  & $\error{69.7}{1.01}{1.08}$ & $\error{68.8}{1.81}{2.68}$ && $\error{70.1}{0.43}{0.40}$ & $\error{69.8}{0.91}{1.30}$ && $\error{70.0}{0.01}{0.01}$ & $\error{69.7}{0.16}{0.16}$ \\
		10\% & $\error{69.9}{0.72}{0.65}$ & $\error{67.6}{2.56}{2.83}$ && $\error{70.1}{0.32}{0.29}$ & $\error{69.3}{1.53}{1.53}$ && $\error{70.0}{0.01}{0.01}$ & $\error{69.6}{0.23}{0.24}$ \\
		\\
		$p(\sigma)$ \\
		0\%  & $\error{0.0}{2.87}{0.00}$ & $\error{0.0}{3.53}{0.00}$ && $\error{0.0}{0.81}{0.00}$ & $\error{0.0}{0.74}{0.00}$ && $\error{0.0}{0.05}{0.00}$ & $\error{1.1}{0.09}{0.07}$ \\
		2\%  & $\error{0.1}{2.40}{0.10}$ & $\error{1.8}{2.79}{1.85}$ && $\error{0.0}{0.80}{0.00}$ & $\error{0.4}{0.97}{0.40}$ && $\error{0.0}{0.05}{0.00}$ & $\error{1.4}{0.15}{0.12}$ \\
		5\%  & $\error{0.0}{2.14}{0.00}$ & $\error{3.8}{2.75}{2.12}$ && $\error{0.0}{0.77}{0.00}$ & $\error{2.3}{1.76}{0.86}$ && $\error{0.0}{0.04}{0.00}$ & $\error{2.1}{0.18}{0.13}$ \\
		10\% & $\error{0.0}{1.58}{0.00}$ & $\error{5.0}{2.18}{2.00}$ && $\error{0.2}{0.46}{0.20}$ & $\error{4.4}{2.24}{1.54}$ && $\error{0.0}{0.04}{0.00}$ & $\error{3.0}{0.39}{0.22}$ \\
		\bottomrule
	\end{tabular}
	\label{tab:mu_sigma_mixture_pop}
\end{table}

We generate a signal using the \eob waveform model and analyze it using the \xphm model.
These waveform models are the two state-of-the-art quasi-circular spin-precessing models incorporating higher-order 
harmonics and used in the analysis of \ac{GW} observations by the \ac{LVK} Collaboration. 
We employ the Fisher matrix method to approximate the posterior distribution using a multivariate Normal distribution 
to estimate measurement accuracies of a binary's parameters.
We use the publicly available Fisher matrix code \gwbench for this purpose. 
Furthermore, we utilize the linear signal approximation to calculate the biases in parameter estimation 
resulting from the differences between the signal model and template model. 
For this calculation, we implement the methodology of \textcite{Dhani:2024jja} for a more faithful representation 
of the systematic biases in parameter estimation. 

In \cref{tab:mu_sigma_mixture_pop}, we report the \ac{MAP} value and 90\% \ac{HPD} interval for $\mu$ and $\sigma$ 
for the different 
detector networks and various fractions of the high-mass spin-precessing population.
We quote the median values across 100 realizations to obtain a noise- and realization-averaged estimate.
In the absence of systematic biases, the \ac{HPD} interval for $\mu$ always encompasses the injected value and $\sigma$ is always 
consistent with zero for all the cases, validating our expectation that in an unbiased measurement the scatter 
of the individual posteriors is consistent with the noise distribution. 
In addition, we find that $H_0$ can be measured at $\sim 3-4\%$ precision with O5 depending on the fraction
of the high-mass spin-precessing population. 
However, we see that even a 5\% contamination with the high-mass spin-precessing population can cause the $\sigma$ 
measurement to exclude zero at 90\% credibility 
in the \HLV and \As networks. 
Interestingly, we find that $\mu$ is consistent with the injected value for all cases in 
\HLV and \As networks with the bias in \XG at a sub-percent level. 
This is partially due to the fact that the uncertainty in $\mu$ is larger 
in the presence of systematic biases, signifying that the 
population hypermodel is able to absorb some of the variation due to systematic biases. This is another advantage of our approach. 
By contrast, we would expect the joint \ac{$H_0$} posterior obtained using the inaccurate model but without the hyper distribution for $H_0$ to be inconsistent with the injected value for such cases. Directly verifying this is difficult in practice 
because one encounters individual event posteriors for which the support does not overlap at machine precision, as described earlier.

Even within the hypermodel framework, the posterior on $\mu$ will tend to a delta function as the number of events increases. A more useful summary of the results is provided by 
the \ac{PPHD} (see \cref{sec:population_diagnostic} for the definition). 
Since \ac{$H_0$} is a constant, in the absence of systematic biases the \ac{PPHD} will approach a delta function as the number of observations increases, just like the posterior on $\mu$. 
However, when systematic biases are present, it will converge to a distribution of finite width,
the latter determined by the fitted variance of the 
population hypermodel. If at least some of the observations are free from systematic biases --- which should be the case --- 
the \ac{PPHD} will encompass the true value.

\textit{Discussion---}
In this letter, we investigated the impact of systematic biases on the inference of $H_0$ due to inaccurate waveform models.
We explored the galaxy catalog method to infer $H_0$ for a \ac{BBH} population
with both planned and future ground-based detector networks.
In the absence of coincident \ac{EM} observations, which seem to be rare, 
dark-siren methods offer the best prospect for standard-siren cosmology. 
With the many tensions in the measurements of cosmological parameters and the increasing evidence for 
dynamical dark energy, alternative and independent measurements are crucial to build consensus and eliminate 
the possibility of systematic errors in any given observation. 

In addition to the putative population of \ac{BBH} mergers observed by the \ac{LVK} Collaboration, 
we studied the impact of a high-mass spin-precessing subpopulation on the $H_0$ estimate. 
Such a population is of astrophysical importance because they are a generic predictions of hierarchical mergers, 
as well as many population synthesis models. 
Furthermore, there are some hints of the existence of such a subpopulation in the GWTC-3 catalog, 
though more observations are needed to infer their properties. 
Its importance in $H_0$ inference is due to the difficulties in modelling such binaries and, therefore, 
the larger systematic biases associated with such mergers. 

We used a different method to detect systematic biases in the inference of $H_0$, based on testing 
the consistency of the variance in the distribution of the individual posteriors with 
the individual measurement errors to identify systematic biases. 
We found that this additional diagnostic is powerful in identifying the presence of systematic biases 
in a population inference which is otherwise not achievable. 
Furthermore, we reported that a small high-mass, spin-precessing subpopulation can be a dominant source of systematic bias even in \HLV and \As.
The effect of systematic biases is even more pronounced for \XG, with $H_0$ estimates from even the GWTC-3 population affected by systematic biases.
Our results exemplify the need for more accurate waveform models for a bright and systematics-free future. 

The main limitation of our study is the reliance on the linear-signal approximation to model the impact of biases. 
Although this method is the most practical way to conduct our study and allows us to 
efficiently analyze thousands of signals spanning the entire parameter space of the \ac{BBH} population, 
we need to better quantify its validity throughout that parameter space.
In the future, we plan to use modern, machine learning based, data analysis tools, such as DINGO~\cite{Dax:2022pxd}, that allow for rapid evaluation of the posteriors. 
Furthermore, our use of two approximate waveform models means that we cannot determine the direction of biases 
in a real \ac{GW} event; we can only highlight the shortcomings of the current models relative to each other. 
We note, however, that the methodology used here is generic and does not rely on knowing the true \ac{GW} signal model. 
Finally, we have only considered a Gaussian model for the fitted population hyper-distribution and alternatives could also be considered. 
However, due to the Central Limit Theorem, we expect that the impact of any kind of systematic will be approximately normally distributed in the limit of a large number of observations, 
so the approach used here should be broadly applicable.

\textit{Acknowledgements---}
The authors are grateful for computational resources provided by the
LIGO Laboratory which National Science Foundation Grants PHY-0757058 and PHY-0823459 support.

\bibliography{references}

\begin{thebibliography}{51}%
\makeatletter
\providecommand \@ifxundefined [1]{%
 \@ifx{#1\undefined}
}%
\providecommand \@ifnum [1]{%
 \ifnum #1\expandafter \@firstoftwo
 \else \expandafter \@secondoftwo
 \fi
}%
\providecommand \@ifx [1]{%
 \ifx #1\expandafter \@firstoftwo
 \else \expandafter \@secondoftwo
 \fi
}%
\providecommand \natexlab [1]{#1}%
\providecommand \enquote  [1]{``#1''}%
\providecommand \bibnamefont  [1]{#1}%
\providecommand \bibfnamefont [1]{#1}%
\providecommand \citenamefont [1]{#1}%
\providecommand \href@noop [0]{\@secondoftwo}%
\providecommand \href [0]{\begingroup \@sanitize@url \@href}%
\providecommand \@href[1]{\@@startlink{#1}\@@href}%
\providecommand \@@href[1]{\endgroup#1\@@endlink}%
\providecommand \@sanitize@url [0]{\catcode `\\12\catcode `\$12\catcode `\&12\catcode `\#12\catcode `\^12\catcode `\_12\catcode `\%12\relax}%
\providecommand \@@startlink[1]{}%
\providecommand \@@endlink[0]{}%
\providecommand \url  [0]{\begingroup\@sanitize@url \@url }%
\providecommand \@url [1]{\endgroup\@href {#1}{\urlprefix }}%
\providecommand \urlprefix  [0]{URL }%
\providecommand \Eprint [0]{\href }%
\providecommand \doibase [0]{https://doi.org/}%
\providecommand \selectlanguage [0]{\@gobble}%
\providecommand \bibinfo  [0]{\@secondoftwo}%
\providecommand \bibfield  [0]{\@secondoftwo}%
\providecommand \translation [1]{[#1]}%
\providecommand \BibitemOpen [0]{}%
\providecommand \bibitemStop [0]{}%
\providecommand \bibitemNoStop [0]{.\EOS\space}%
\providecommand \EOS [0]{\spacefactor3000\relax}%
\providecommand \BibitemShut  [1]{\csname bibitem#1\endcsname}%
\let\auto@bib@innerbib\@empty
\bibitem [{\citenamefont {Abbott}\ \emph {et~al.}(2017{\natexlab{a}})\citenamefont {Abbott} \emph {et~al.}}]{LIGOScientific:2017adf}%
  \BibitemOpen
  \bibfield  {author} {\bibinfo {author} {\bibfnamefont {B.~P.}\ \bibnamefont {Abbott}} \emph {et~al.} (\bibinfo {collaboration} {LIGO Scientific, Virgo, 1M2H, Dark Energy Camera GW-E, DES, DLT40, Las Cumbres Observatory, VINROUGE, MASTER}),\ }\bibfield  {title} {\bibinfo {title} {{A gravitational-wave standard siren measurement of the Hubble constant}},\ }\href {https://doi.org/10.1038/nature24471} {\bibfield  {journal} {\bibinfo  {journal} {Nature}\ }\textbf {\bibinfo {volume} {551}},\ \bibinfo {pages} {85} (\bibinfo {year} {2017}{\natexlab{a}})},\ \Eprint {https://arxiv.org/abs/1710.05835} {arXiv:1710.05835 [astro-ph.CO]} \BibitemShut {NoStop}%
\bibitem [{\citenamefont {Abbott}\ \emph {et~al.}(2023{\natexlab{a}})\citenamefont {Abbott} \emph {et~al.}}]{LIGOScientific:2021aug}%
  \BibitemOpen
  \bibfield  {author} {\bibinfo {author} {\bibfnamefont {R.}~\bibnamefont {Abbott}} \emph {et~al.} (\bibinfo {collaboration} {LIGO Scientific, Virgo, KAGRA}),\ }\bibfield  {title} {\bibinfo {title} {{Constraints on the Cosmic Expansion History from GWTC\textendash{}3}},\ }\href {https://doi.org/10.3847/1538-4357/ac74bb} {\bibfield  {journal} {\bibinfo  {journal} {Astrophys. J.}\ }\textbf {\bibinfo {volume} {949}},\ \bibinfo {pages} {76} (\bibinfo {year} {2023}{\natexlab{a}})},\ \Eprint {https://arxiv.org/abs/2111.03604} {arXiv:2111.03604 [astro-ph.CO]} \BibitemShut {NoStop}%
\bibitem [{\citenamefont {{Schutz}}(1986)}]{1986Natur.323..310S}%
  \BibitemOpen
  \bibfield  {author} {\bibinfo {author} {\bibfnamefont {B.~F.}\ \bibnamefont {{Schutz}}},\ }\bibfield  {title} {\bibinfo {title} {{Determining the Hubble constant from gravitational wave observations}},\ }\href {https://doi.org/10.1038/323310a0} {\bibfield  {journal} {\bibinfo  {journal} {\nat}\ }\textbf {\bibinfo {volume} {323}},\ \bibinfo {pages} {310} (\bibinfo {year} {1986})}\BibitemShut {NoStop}%
\bibitem [{\citenamefont {Fishbach}\ \emph {et~al.}(2019)\citenamefont {Fishbach} \emph {et~al.}}]{LIGOScientific:2018gmd}%
  \BibitemOpen
  \bibfield  {author} {\bibinfo {author} {\bibfnamefont {M.}~\bibnamefont {Fishbach}} \emph {et~al.} (\bibinfo {collaboration} {LIGO Scientific, Virgo}),\ }\bibfield  {title} {\bibinfo {title} {{A Standard Siren Measurement of the Hubble Constant from GW170817 without the Electromagnetic Counterpart}},\ }\href {https://doi.org/10.3847/2041-8213/aaf96e} {\bibfield  {journal} {\bibinfo  {journal} {Astrophys. J. Lett.}\ }\textbf {\bibinfo {volume} {871}},\ \bibinfo {pages} {L13} (\bibinfo {year} {2019})},\ \Eprint {https://arxiv.org/abs/1807.05667} {arXiv:1807.05667 [astro-ph.CO]} \BibitemShut {NoStop}%
\bibitem [{\citenamefont {Del~Pozzo}(2012)}]{DelPozzo:2011vcw}%
  \BibitemOpen
  \bibfield  {author} {\bibinfo {author} {\bibfnamefont {W.}~\bibnamefont {Del~Pozzo}},\ }\bibfield  {title} {\bibinfo {title} {{Inference of the cosmological parameters from gravitational waves: application to second generation interferometers}},\ }\href {https://doi.org/10.1103/PhysRevD.86.043011} {\bibfield  {journal} {\bibinfo  {journal} {Phys. Rev. D}\ }\textbf {\bibinfo {volume} {86}},\ \bibinfo {pages} {043011} (\bibinfo {year} {2012})},\ \Eprint {https://arxiv.org/abs/1108.1317} {arXiv:1108.1317 [astro-ph.CO]} \BibitemShut {NoStop}%
\bibitem [{\citenamefont {Borhanian}\ \emph {et~al.}(2020)\citenamefont {Borhanian}, \citenamefont {Dhani}, \citenamefont {Gupta}, \citenamefont {Arun},\ and\ \citenamefont {Sathyaprakash}}]{Borhanian:2020vyr}%
  \BibitemOpen
  \bibfield  {author} {\bibinfo {author} {\bibfnamefont {S.}~\bibnamefont {Borhanian}}, \bibinfo {author} {\bibfnamefont {A.}~\bibnamefont {Dhani}}, \bibinfo {author} {\bibfnamefont {A.}~\bibnamefont {Gupta}}, \bibinfo {author} {\bibfnamefont {K.~G.}\ \bibnamefont {Arun}},\ and\ \bibinfo {author} {\bibfnamefont {B.~S.}\ \bibnamefont {Sathyaprakash}},\ }\bibfield  {title} {\bibinfo {title} {{Dark Sirens to Resolve the Hubble\textendash{}Lema\^\i{}tre Tension}},\ }\href {https://doi.org/10.3847/2041-8213/abcaf5} {\bibfield  {journal} {\bibinfo  {journal} {Astrophys. J. Lett.}\ }\textbf {\bibinfo {volume} {905}},\ \bibinfo {pages} {L28} (\bibinfo {year} {2020})},\ \Eprint {https://arxiv.org/abs/2007.02883} {arXiv:2007.02883 [astro-ph.CO]} \BibitemShut {NoStop}%
\bibitem [{\citenamefont {Mukherjee}\ \emph {et~al.}(2021)\citenamefont {Mukherjee}, \citenamefont {Wandelt}, \citenamefont {Nissanke},\ and\ \citenamefont {Silvestri}}]{PhysRevD.103.043520}%
  \BibitemOpen
  \bibfield  {author} {\bibinfo {author} {\bibfnamefont {S.}~\bibnamefont {Mukherjee}}, \bibinfo {author} {\bibfnamefont {B.~D.}\ \bibnamefont {Wandelt}}, \bibinfo {author} {\bibfnamefont {S.~M.}\ \bibnamefont {Nissanke}},\ and\ \bibinfo {author} {\bibfnamefont {A.}~\bibnamefont {Silvestri}},\ }\bibfield  {title} {\bibinfo {title} {Accurate precision cosmology with redshift unknown gravitational wave sources},\ }\href {https://doi.org/10.1103/PhysRevD.103.043520} {\bibfield  {journal} {\bibinfo  {journal} {Phys. Rev. D}\ }\textbf {\bibinfo {volume} {103}},\ \bibinfo {pages} {043520} (\bibinfo {year} {2021})}\BibitemShut {NoStop}%
\bibitem [{\citenamefont {Mukherjee}\ \emph {et~al.}(2024)\citenamefont {Mukherjee}, \citenamefont {Krolewski}, \citenamefont {Wandelt},\ and\ \citenamefont {Silk}}]{Mukherjee:2022afz}%
  \BibitemOpen
  \bibfield  {author} {\bibinfo {author} {\bibfnamefont {S.}~\bibnamefont {Mukherjee}}, \bibinfo {author} {\bibfnamefont {A.}~\bibnamefont {Krolewski}}, \bibinfo {author} {\bibfnamefont {B.~D.}\ \bibnamefont {Wandelt}},\ and\ \bibinfo {author} {\bibfnamefont {J.}~\bibnamefont {Silk}},\ }\bibfield  {title} {\bibinfo {title} {{Cross-correlating dark sirens and galaxies: constraints on $H_0$ from GWTC-3 of LIGO-Virgo-KAGRA}},\ }\href {https://doi.org/10.3847/1538-4357/ad7d90} {\bibfield  {journal} {\bibinfo  {journal} {Astrophys. J.}\ }\textbf {\bibinfo {volume} {975}},\ \bibinfo {pages} {189} (\bibinfo {year} {2024})},\ \Eprint {https://arxiv.org/abs/2203.03643} {arXiv:2203.03643 [astro-ph.CO]} \BibitemShut {NoStop}%
\bibitem [{\citenamefont {Chernoff}\ and\ \citenamefont {Finn}(1993)}]{Chernoff:1993th}%
  \BibitemOpen
  \bibfield  {author} {\bibinfo {author} {\bibfnamefont {D.~F.}\ \bibnamefont {Chernoff}}\ and\ \bibinfo {author} {\bibfnamefont {L.~S.}\ \bibnamefont {Finn}},\ }\bibfield  {title} {\bibinfo {title} {{Gravitational radiation, inspiraling binaries, and cosmology}},\ }\href {https://doi.org/10.1086/186898} {\bibfield  {journal} {\bibinfo  {journal} {Astrophys. J. Lett.}\ }\textbf {\bibinfo {volume} {411}},\ \bibinfo {pages} {L5} (\bibinfo {year} {1993})},\ \Eprint {https://arxiv.org/abs/gr-qc/9304020} {arXiv:gr-qc/9304020} \BibitemShut {NoStop}%
\bibitem [{\citenamefont {Taylor}\ and\ \citenamefont {Gair}(2012)}]{Taylor:2012db}%
  \BibitemOpen
  \bibfield  {author} {\bibinfo {author} {\bibfnamefont {S.~R.}\ \bibnamefont {Taylor}}\ and\ \bibinfo {author} {\bibfnamefont {J.~R.}\ \bibnamefont {Gair}},\ }\bibfield  {title} {\bibinfo {title} {{Cosmology with the lights off: standard sirens in the Einstein Telescope era}},\ }\href {https://doi.org/10.1103/PhysRevD.86.023502} {\bibfield  {journal} {\bibinfo  {journal} {Phys. Rev. D}\ }\textbf {\bibinfo {volume} {86}},\ \bibinfo {pages} {023502} (\bibinfo {year} {2012})},\ \Eprint {https://arxiv.org/abs/1204.6739} {arXiv:1204.6739 [astro-ph.CO]} \BibitemShut {NoStop}%
\bibitem [{\citenamefont {Farr}\ \emph {et~al.}(2019)\citenamefont {Farr}, \citenamefont {Fishbach}, \citenamefont {Ye},\ and\ \citenamefont {Holz}}]{Farr:2019twy}%
  \BibitemOpen
  \bibfield  {author} {\bibinfo {author} {\bibfnamefont {W.~M.}\ \bibnamefont {Farr}}, \bibinfo {author} {\bibfnamefont {M.}~\bibnamefont {Fishbach}}, \bibinfo {author} {\bibfnamefont {J.}~\bibnamefont {Ye}},\ and\ \bibinfo {author} {\bibfnamefont {D.}~\bibnamefont {Holz}},\ }\bibfield  {title} {\bibinfo {title} {{A Future Percent-Level Measurement of the Hubble Expansion at Redshift 0.8 With Advanced LIGO}},\ }\href {https://doi.org/10.3847/2041-8213/ab4284} {\bibfield  {journal} {\bibinfo  {journal} {Astrophys. J. Lett.}\ }\textbf {\bibinfo {volume} {883}},\ \bibinfo {pages} {L42} (\bibinfo {year} {2019})},\ \Eprint {https://arxiv.org/abs/1908.09084} {arXiv:1908.09084 [astro-ph.CO]} \BibitemShut {NoStop}%
\bibitem [{\citenamefont {Ezquiaga}\ and\ \citenamefont {Holz}(2022)}]{Ezquiaga:2022zkx}%
  \BibitemOpen
  \bibfield  {author} {\bibinfo {author} {\bibfnamefont {J.~M.}\ \bibnamefont {Ezquiaga}}\ and\ \bibinfo {author} {\bibfnamefont {D.~E.}\ \bibnamefont {Holz}},\ }\bibfield  {title} {\bibinfo {title} {{Spectral Sirens: Cosmology from the Full Mass Distribution of Compact Binaries}},\ }\href {https://doi.org/10.1103/PhysRevLett.129.061102} {\bibfield  {journal} {\bibinfo  {journal} {Phys. Rev. Lett.}\ }\textbf {\bibinfo {volume} {129}},\ \bibinfo {pages} {061102} (\bibinfo {year} {2022})},\ \Eprint {https://arxiv.org/abs/2202.08240} {arXiv:2202.08240 [astro-ph.CO]} \BibitemShut {NoStop}%
\bibitem [{\citenamefont {Tong}\ \emph {et~al.}(2025)\citenamefont {Tong}, \citenamefont {Fishbach},\ and\ \citenamefont {Thrane}}]{Tong:2025xvd}%
  \BibitemOpen
  \bibfield  {author} {\bibinfo {author} {\bibfnamefont {H.}~\bibnamefont {Tong}}, \bibinfo {author} {\bibfnamefont {M.}~\bibnamefont {Fishbach}},\ and\ \bibinfo {author} {\bibfnamefont {E.}~\bibnamefont {Thrane}},\ }\bibfield  {title} {\bibinfo {title} {{Spinning Spectral Sirens: Robust Cosmological Measurement Using Mass{\textendash}Spin Correlations in the Binary Black Hole Population}},\ }\href {https://doi.org/10.3847/1538-4357/adcec5} {\bibfield  {journal} {\bibinfo  {journal} {Astrophys. J.}\ }\textbf {\bibinfo {volume} {985}},\ \bibinfo {pages} {220} (\bibinfo {year} {2025})},\ \Eprint {https://arxiv.org/abs/2502.10780} {arXiv:2502.10780 [astro-ph.CO]} \BibitemShut {NoStop}%
\bibitem [{\citenamefont {Cousins}\ \emph {et~al.}(2025)\citenamefont {Cousins}, \citenamefont {Schumacher}, \citenamefont {Chung}, \citenamefont {Talbot}, \citenamefont {Callister}, \citenamefont {Holz},\ and\ \citenamefont {Yunes}}]{Cousins:2025bas}%
  \BibitemOpen
  \bibfield  {author} {\bibinfo {author} {\bibfnamefont {B.}~\bibnamefont {Cousins}}, \bibinfo {author} {\bibfnamefont {K.}~\bibnamefont {Schumacher}}, \bibinfo {author} {\bibfnamefont {A.~K.-W.}\ \bibnamefont {Chung}}, \bibinfo {author} {\bibfnamefont {C.}~\bibnamefont {Talbot}}, \bibinfo {author} {\bibfnamefont {T.}~\bibnamefont {Callister}}, \bibinfo {author} {\bibfnamefont {D.~E.}\ \bibnamefont {Holz}},\ and\ \bibinfo {author} {\bibfnamefont {N.}~\bibnamefont {Yunes}},\ }\bibfield  {title} {\bibinfo {title} {{The Stochastic Siren: Astrophysical Gravitational-Wave Background Measurements of the Hubble Constant}},\ }\href@noop {} {\  (\bibinfo {year} {2025})},\ \Eprint {https://arxiv.org/abs/2503.01997} {arXiv:2503.01997 [astro-ph.CO]} \BibitemShut {NoStop}%
\bibitem [{\citenamefont {Messenger}\ and\ \citenamefont {Read}(2012)}]{Messenger:2011gi}%
  \BibitemOpen
  \bibfield  {author} {\bibinfo {author} {\bibfnamefont {C.}~\bibnamefont {Messenger}}\ and\ \bibinfo {author} {\bibfnamefont {J.}~\bibnamefont {Read}},\ }\bibfield  {title} {\bibinfo {title} {{Measuring a cosmological distance-redshift relationship using only gravitational wave observations of binary neutron star coalescences}},\ }\href {https://doi.org/10.1103/PhysRevLett.108.091101} {\bibfield  {journal} {\bibinfo  {journal} {Phys. Rev. Lett.}\ }\textbf {\bibinfo {volume} {108}},\ \bibinfo {pages} {091101} (\bibinfo {year} {2012})},\ \Eprint {https://arxiv.org/abs/1107.5725} {arXiv:1107.5725 [gr-qc]} \BibitemShut {NoStop}%
\bibitem [{\citenamefont {Messenger}\ \emph {et~al.}(2014)\citenamefont {Messenger}, \citenamefont {Takami}, \citenamefont {Gossan}, \citenamefont {Rezzolla},\ and\ \citenamefont {Sathyaprakash}}]{Messenger:2013fya}%
  \BibitemOpen
  \bibfield  {author} {\bibinfo {author} {\bibfnamefont {C.}~\bibnamefont {Messenger}}, \bibinfo {author} {\bibfnamefont {K.}~\bibnamefont {Takami}}, \bibinfo {author} {\bibfnamefont {S.}~\bibnamefont {Gossan}}, \bibinfo {author} {\bibfnamefont {L.}~\bibnamefont {Rezzolla}},\ and\ \bibinfo {author} {\bibfnamefont {B.~S.}\ \bibnamefont {Sathyaprakash}},\ }\bibfield  {title} {\bibinfo {title} {{Source Redshifts from Gravitational-Wave Observations of Binary Neutron Star Mergers}},\ }\href {https://doi.org/10.1103/PhysRevX.4.041004} {\bibfield  {journal} {\bibinfo  {journal} {Phys. Rev. X}\ }\textbf {\bibinfo {volume} {4}},\ \bibinfo {pages} {041004} (\bibinfo {year} {2014})},\ \Eprint {https://arxiv.org/abs/1312.1862} {arXiv:1312.1862 [gr-qc]} \BibitemShut {NoStop}%
\bibitem [{\citenamefont {Li}\ \emph {et~al.}(2015)\citenamefont {Li}, \citenamefont {Del~Pozzo},\ and\ \citenamefont {Messenger}}]{Li:2013via}%
  \BibitemOpen
  \bibfield  {author} {\bibinfo {author} {\bibfnamefont {T.~G.~F.}\ \bibnamefont {Li}}, \bibinfo {author} {\bibfnamefont {W.}~\bibnamefont {Del~Pozzo}},\ and\ \bibinfo {author} {\bibfnamefont {C.}~\bibnamefont {Messenger}},\ }\bibfield  {title} {\bibinfo {title} {{Measuring the redshift of standard sirens using the neutron star deformability}},\ }in\ \href {https://doi.org/10.1142/9789814623995_0346} {\emph {\bibinfo {booktitle} {{13th Marcel Grossmann Meeting on Recent Developments in Theoretical and Experimental General Relativity, Astrophysics, and Relativistic Field Theories}}}}\ (\bibinfo {year} {2015})\ pp.\ \bibinfo {pages} {2019--2021},\ \Eprint {https://arxiv.org/abs/1303.0855} {arXiv:1303.0855 [gr-qc]} \BibitemShut {NoStop}%
\bibitem [{\citenamefont {Dhani}\ \emph {et~al.}(2022)\citenamefont {Dhani}, \citenamefont {Borhanian}, \citenamefont {Gupta},\ and\ \citenamefont {Sathyaprakash}}]{Dhani:2022ulg}%
  \BibitemOpen
  \bibfield  {author} {\bibinfo {author} {\bibfnamefont {A.}~\bibnamefont {Dhani}}, \bibinfo {author} {\bibfnamefont {S.}~\bibnamefont {Borhanian}}, \bibinfo {author} {\bibfnamefont {A.}~\bibnamefont {Gupta}},\ and\ \bibinfo {author} {\bibfnamefont {B.}~\bibnamefont {Sathyaprakash}},\ }\bibfield  {title} {\bibinfo {title} {{Cosmography with bright and Love sirens}},\ }\Eprint {https://arxiv.org/abs/2212.13183} {arXiv:2212.13183 [gr-qc]}  (\bibinfo {year} {2022})\BibitemShut {NoStop}%
\bibitem [{\citenamefont {Aasi}\ \emph {et~al.}(2015)\citenamefont {Aasi} \emph {et~al.}}]{LIGOScientific:2014pky}%
  \BibitemOpen
  \bibfield  {author} {\bibinfo {author} {\bibfnamefont {J.}~\bibnamefont {Aasi}} \emph {et~al.} (\bibinfo {collaboration} {LIGO Scientific}),\ }\bibfield  {title} {\bibinfo {title} {{Advanced LIGO}},\ }\href {https://doi.org/10.1088/0264-9381/32/7/074001} {\bibfield  {journal} {\bibinfo  {journal} {Class. Quant. Grav.}\ }\textbf {\bibinfo {volume} {32}},\ \bibinfo {pages} {074001} (\bibinfo {year} {2015})},\ \Eprint {https://arxiv.org/abs/1411.4547} {arXiv:1411.4547 [gr-qc]} \BibitemShut {NoStop}%
\bibitem [{\citenamefont {Acernese}\ \emph {et~al.}(2015)\citenamefont {Acernese} \emph {et~al.}}]{VIRGO:2014yos}%
  \BibitemOpen
  \bibfield  {author} {\bibinfo {author} {\bibfnamefont {F.}~\bibnamefont {Acernese}} \emph {et~al.} (\bibinfo {collaboration} {VIRGO}),\ }\bibfield  {title} {\bibinfo {title} {{Advanced Virgo: a second-generation interferometric gravitational wave detector}},\ }\href {https://doi.org/10.1088/0264-9381/32/2/024001} {\bibfield  {journal} {\bibinfo  {journal} {Class. Quant. Grav.}\ }\textbf {\bibinfo {volume} {32}},\ \bibinfo {pages} {024001} (\bibinfo {year} {2015})},\ \Eprint {https://arxiv.org/abs/1408.3978} {arXiv:1408.3978 [gr-qc]} \BibitemShut {NoStop}%
\bibitem [{\citenamefont {Akutsu}\ \emph {et~al.}(2021)\citenamefont {Akutsu} \emph {et~al.}}]{KAGRA:2020tym}%
  \BibitemOpen
  \bibfield  {author} {\bibinfo {author} {\bibfnamefont {T.}~\bibnamefont {Akutsu}} \emph {et~al.} (\bibinfo {collaboration} {KAGRA}),\ }\bibfield  {title} {\bibinfo {title} {{Overview of KAGRA: Detector design and construction history}},\ }\href {https://doi.org/10.1093/ptep/ptaa125} {\bibfield  {journal} {\bibinfo  {journal} {PTEP}\ }\textbf {\bibinfo {volume} {2021}},\ \bibinfo {pages} {05A101} (\bibinfo {year} {2021})},\ \Eprint {https://arxiv.org/abs/2005.05574} {arXiv:2005.05574 [physics.ins-det]} \BibitemShut {NoStop}%
\bibitem [{\citenamefont {Borhanian}\ and\ \citenamefont {Sathyaprakash}(2024)}]{Borhanian:2022czq}%
  \BibitemOpen
  \bibfield  {author} {\bibinfo {author} {\bibfnamefont {S.}~\bibnamefont {Borhanian}}\ and\ \bibinfo {author} {\bibfnamefont {B.~S.}\ \bibnamefont {Sathyaprakash}},\ }\bibfield  {title} {\bibinfo {title} {{Listening to the Universe with next generation ground-based gravitational-wave detectors}},\ }\href {https://doi.org/10.1103/PhysRevD.110.083040} {\bibfield  {journal} {\bibinfo  {journal} {Phys. Rev. D}\ }\textbf {\bibinfo {volume} {110}},\ \bibinfo {pages} {083040} (\bibinfo {year} {2024})},\ \Eprint {https://arxiv.org/abs/2202.11048} {arXiv:2202.11048 [gr-qc]} \BibitemShut {NoStop}%
\bibitem [{\citenamefont {Reitze}\ \emph {et~al.}(2019)\citenamefont {Reitze} \emph {et~al.}}]{Reitze:2019iox}%
  \BibitemOpen
  \bibfield  {author} {\bibinfo {author} {\bibfnamefont {D.}~\bibnamefont {Reitze}} \emph {et~al.},\ }\bibfield  {title} {\bibinfo {title} {{Cosmic Explorer: The U.S. Contribution to Gravitational-Wave Astronomy beyond LIGO}},\ }\href@noop {} {\bibfield  {journal} {\bibinfo  {journal} {Bull. Am. Astron. Soc.}\ }\textbf {\bibinfo {volume} {51}},\ \bibinfo {pages} {035} (\bibinfo {year} {2019})},\ \Eprint {https://arxiv.org/abs/1907.04833} {arXiv:1907.04833 [astro-ph.IM]} \BibitemShut {NoStop}%
\bibitem [{\citenamefont {Evans}\ \emph {et~al.}(2021)\citenamefont {Evans} \emph {et~al.}}]{Evans:2021gyd}%
  \BibitemOpen
  \bibfield  {author} {\bibinfo {author} {\bibfnamefont {M.}~\bibnamefont {Evans}} \emph {et~al.},\ }\bibfield  {title} {\bibinfo {title} {{A Horizon Study for Cosmic Explorer: Science, Observatories, and Community}},\ }\Eprint {https://arxiv.org/abs/2109.09882} {arXiv:2109.09882 [astro-ph.IM]}  (\bibinfo {year} {2021})\BibitemShut {NoStop}%
\bibitem [{\citenamefont {Srivastava}\ \emph {et~al.}(2022)\citenamefont {Srivastava}, \citenamefont {Davis}, \citenamefont {Kuns}, \citenamefont {Landry}, \citenamefont {Ballmer}, \citenamefont {Evans}, \citenamefont {Hall}, \citenamefont {Read},\ and\ \citenamefont {Sathyaprakash}}]{Srivastava:2022slt}%
  \BibitemOpen
  \bibfield  {author} {\bibinfo {author} {\bibfnamefont {V.}~\bibnamefont {Srivastava}}, \bibinfo {author} {\bibfnamefont {D.}~\bibnamefont {Davis}}, \bibinfo {author} {\bibfnamefont {K.}~\bibnamefont {Kuns}}, \bibinfo {author} {\bibfnamefont {P.}~\bibnamefont {Landry}}, \bibinfo {author} {\bibfnamefont {S.}~\bibnamefont {Ballmer}}, \bibinfo {author} {\bibfnamefont {M.}~\bibnamefont {Evans}}, \bibinfo {author} {\bibfnamefont {E.~D.}\ \bibnamefont {Hall}}, \bibinfo {author} {\bibfnamefont {J.}~\bibnamefont {Read}},\ and\ \bibinfo {author} {\bibfnamefont {B.~S.}\ \bibnamefont {Sathyaprakash}},\ }\bibfield  {title} {\bibinfo {title} {{Science-driven Tunable Design of Cosmic Explorer Detectors}},\ }\href {https://doi.org/10.3847/1538-4357/ac5f04} {\bibfield  {journal} {\bibinfo  {journal} {Astrophys. J.}\ }\textbf {\bibinfo {volume} {931}},\ \bibinfo {pages} {22} (\bibinfo {year} {2022})},\ \Eprint {https://arxiv.org/abs/2201.10668} {arXiv:2201.10668 [gr-qc]} \BibitemShut {NoStop}%
\bibitem [{\citenamefont {Punturo}\ \emph {et~al.}(2010)\citenamefont {Punturo} \emph {et~al.}}]{Punturo:2010zz}%
  \BibitemOpen
  \bibfield  {author} {\bibinfo {author} {\bibfnamefont {M.}~\bibnamefont {Punturo}} \emph {et~al.},\ }\bibfield  {title} {\bibinfo {title} {{The Einstein Telescope: A third-generation gravitational wave observatory}},\ }\href {https://doi.org/10.1088/0264-9381/27/19/194002} {\bibfield  {journal} {\bibinfo  {journal} {Class. Quant. Grav.}\ }\textbf {\bibinfo {volume} {27}},\ \bibinfo {pages} {194002} (\bibinfo {year} {2010})}\BibitemShut {NoStop}%
\bibitem [{\citenamefont {Maggiore}\ \emph {et~al.}(2020)\citenamefont {Maggiore} \emph {et~al.}}]{Maggiore:2019uih}%
  \BibitemOpen
  \bibfield  {author} {\bibinfo {author} {\bibfnamefont {M.}~\bibnamefont {Maggiore}} \emph {et~al.},\ }\bibfield  {title} {\bibinfo {title} {{Science Case for the Einstein Telescope}},\ }\href {https://doi.org/10.1088/1475-7516/2020/03/050} {\bibfield  {journal} {\bibinfo  {journal} {JCAP}\ }\textbf {\bibinfo {volume} {03}},\ \bibinfo {pages} {050}},\ \Eprint {https://arxiv.org/abs/1912.02622} {arXiv:1912.02622 [astro-ph.CO]} \BibitemShut {NoStop}%
\bibitem [{\citenamefont {Branchesi}\ \emph {et~al.}(2023)\citenamefont {Branchesi} \emph {et~al.}}]{Branchesi:2023mws}%
  \BibitemOpen
  \bibfield  {author} {\bibinfo {author} {\bibfnamefont {M.}~\bibnamefont {Branchesi}} \emph {et~al.},\ }\bibfield  {title} {\bibinfo {title} {{Science with the Einstein Telescope: a comparison of different designs}},\ }\href {https://doi.org/10.1088/1475-7516/2023/07/068} {\bibfield  {journal} {\bibinfo  {journal} {JCAP}\ }\textbf {\bibinfo {volume} {07}},\ \bibinfo {pages} {068}},\ \Eprint {https://arxiv.org/abs/2303.15923} {arXiv:2303.15923 [gr-qc]} \BibitemShut {NoStop}%
\bibitem [{\citenamefont {Muttoni}\ \emph {et~al.}(2023)\citenamefont {Muttoni}, \citenamefont {Laghi}, \citenamefont {Tamanini}, \citenamefont {Marsat},\ and\ \citenamefont {Izquierdo-Villalba}}]{Muttoni:2023prw}%
  \BibitemOpen
  \bibfield  {author} {\bibinfo {author} {\bibfnamefont {N.}~\bibnamefont {Muttoni}}, \bibinfo {author} {\bibfnamefont {D.}~\bibnamefont {Laghi}}, \bibinfo {author} {\bibfnamefont {N.}~\bibnamefont {Tamanini}}, \bibinfo {author} {\bibfnamefont {S.}~\bibnamefont {Marsat}},\ and\ \bibinfo {author} {\bibfnamefont {D.}~\bibnamefont {Izquierdo-Villalba}},\ }\bibfield  {title} {\bibinfo {title} {{Dark siren cosmology with binary black holes in the era of third-generation gravitational wave detectors}},\ }\href {https://doi.org/10.1103/PhysRevD.108.043543} {\bibfield  {journal} {\bibinfo  {journal} {Phys. Rev. D}\ }\textbf {\bibinfo {volume} {108}},\ \bibinfo {pages} {043543} (\bibinfo {year} {2023})},\ \Eprint {https://arxiv.org/abs/2303.10693} {arXiv:2303.10693 [astro-ph.CO]} \BibitemShut {NoStop}%
\bibitem [{\citenamefont {Aghanim}\ \emph {et~al.}(2020)\citenamefont {Aghanim} \emph {et~al.}}]{Planck:2018vyg}%
  \BibitemOpen
  \bibfield  {author} {\bibinfo {author} {\bibfnamefont {N.}~\bibnamefont {Aghanim}} \emph {et~al.} (\bibinfo {collaboration} {Planck}),\ }\bibfield  {title} {\bibinfo {title} {{Planck 2018 results. VI. Cosmological parameters}},\ }\href {https://doi.org/10.1051/0004-6361/201833910} {\bibfield  {journal} {\bibinfo  {journal} {Astron. Astrophys.}\ }\textbf {\bibinfo {volume} {641}},\ \bibinfo {pages} {A6} (\bibinfo {year} {2020})},\ \bibinfo {note} {[Erratum: Astron.Astrophys. 652, C4 (2021)]},\ \Eprint {https://arxiv.org/abs/1807.06209} {arXiv:1807.06209 [astro-ph.CO]} \BibitemShut {NoStop}%
\bibitem [{\citenamefont {Riess}\ \emph {et~al.}(2022)\citenamefont {Riess} \emph {et~al.}}]{Riess:2021jrx}%
  \BibitemOpen
  \bibfield  {author} {\bibinfo {author} {\bibfnamefont {A.~G.}\ \bibnamefont {Riess}} \emph {et~al.},\ }\bibfield  {title} {\bibinfo {title} {{A Comprehensive Measurement of the Local Value of the Hubble Constant with 1 km s$^{-1}$ Mpc$^{-1}$ Uncertainty from the Hubble Space Telescope and the SH0ES Team}},\ }\href {https://doi.org/10.3847/2041-8213/ac5c5b} {\bibfield  {journal} {\bibinfo  {journal} {Astrophys. J. Lett.}\ }\textbf {\bibinfo {volume} {934}},\ \bibinfo {pages} {L7} (\bibinfo {year} {2022})},\ \Eprint {https://arxiv.org/abs/2112.04510} {arXiv:2112.04510 [astro-ph.CO]} \BibitemShut {NoStop}%
\bibitem [{\citenamefont {Owen}\ \emph {et~al.}(2023)\citenamefont {Owen}, \citenamefont {Haster}, \citenamefont {Perkins}, \citenamefont {Cornish},\ and\ \citenamefont {Yunes}}]{Owen:2023mid}%
  \BibitemOpen
  \bibfield  {author} {\bibinfo {author} {\bibfnamefont {C.~B.}\ \bibnamefont {Owen}}, \bibinfo {author} {\bibfnamefont {C.-J.}\ \bibnamefont {Haster}}, \bibinfo {author} {\bibfnamefont {S.}~\bibnamefont {Perkins}}, \bibinfo {author} {\bibfnamefont {N.~J.}\ \bibnamefont {Cornish}},\ and\ \bibinfo {author} {\bibfnamefont {N.}~\bibnamefont {Yunes}},\ }\bibfield  {title} {\bibinfo {title} {{Waveform accuracy and systematic uncertainties in current gravitational wave observations}},\ }\href {https://doi.org/10.1103/PhysRevD.108.044018} {\bibfield  {journal} {\bibinfo  {journal} {Phys. Rev. D}\ }\textbf {\bibinfo {volume} {108}},\ \bibinfo {pages} {044018} (\bibinfo {year} {2023})},\ \Eprint {https://arxiv.org/abs/2301.11941} {arXiv:2301.11941 [gr-qc]} \BibitemShut {NoStop}%
\bibitem [{\citenamefont {Dhani}\ \emph {et~al.}(2024)\citenamefont {Dhani}, \citenamefont {V\"olkel}, \citenamefont {Buonanno}, \citenamefont {Estelles}, \citenamefont {Gair}, \citenamefont {Pfeiffer}, \citenamefont {Pompili},\ and\ \citenamefont {Toubiana}}]{Dhani:2024jja}%
  \BibitemOpen
  \bibfield  {author} {\bibinfo {author} {\bibfnamefont {A.}~\bibnamefont {Dhani}}, \bibinfo {author} {\bibfnamefont {S.}~\bibnamefont {V\"olkel}}, \bibinfo {author} {\bibfnamefont {A.}~\bibnamefont {Buonanno}}, \bibinfo {author} {\bibfnamefont {H.}~\bibnamefont {Estelles}}, \bibinfo {author} {\bibfnamefont {J.}~\bibnamefont {Gair}}, \bibinfo {author} {\bibfnamefont {H.~P.}\ \bibnamefont {Pfeiffer}}, \bibinfo {author} {\bibfnamefont {L.}~\bibnamefont {Pompili}},\ and\ \bibinfo {author} {\bibfnamefont {A.}~\bibnamefont {Toubiana}},\ }\bibfield  {title} {\bibinfo {title} {{Systematic Biases in Estimating the Properties of Black Holes Due to Inaccurate Gravitational-Wave Models}},\ }\href@noop {} {\  (\bibinfo {year} {2024})},\ \Eprint {https://arxiv.org/abs/2404.05811} {arXiv:2404.05811 [gr-qc]} \BibitemShut {NoStop}%
\bibitem [{\citenamefont {Kapil}\ \emph {et~al.}(2024)\citenamefont {Kapil}, \citenamefont {Reali}, \citenamefont {Cotesta},\ and\ \citenamefont {Berti}}]{Kapil:2024zdn}%
  \BibitemOpen
  \bibfield  {author} {\bibinfo {author} {\bibfnamefont {V.}~\bibnamefont {Kapil}}, \bibinfo {author} {\bibfnamefont {L.}~\bibnamefont {Reali}}, \bibinfo {author} {\bibfnamefont {R.}~\bibnamefont {Cotesta}},\ and\ \bibinfo {author} {\bibfnamefont {E.}~\bibnamefont {Berti}},\ }\bibfield  {title} {\bibinfo {title} {{Systematic bias from waveform modeling for binary black hole populations in next-generation gravitational wave detectors}},\ }\href {https://doi.org/10.1103/PhysRevD.109.104043} {\bibfield  {journal} {\bibinfo  {journal} {Phys. Rev. D}\ }\textbf {\bibinfo {volume} {109}},\ \bibinfo {pages} {104043} (\bibinfo {year} {2024})},\ \Eprint {https://arxiv.org/abs/2404.00090} {arXiv:2404.00090 [gr-qc]} \BibitemShut {NoStop}%
\bibitem [{\citenamefont {P\"urrer}\ and\ \citenamefont {Haster}(2020)}]{Purrer:2019jcp}%
  \BibitemOpen
  \bibfield  {author} {\bibinfo {author} {\bibfnamefont {M.}~\bibnamefont {P\"urrer}}\ and\ \bibinfo {author} {\bibfnamefont {C.-J.}\ \bibnamefont {Haster}},\ }\bibfield  {title} {\bibinfo {title} {{Gravitational waveform accuracy requirements for future ground-based detectors}},\ }\href {https://doi.org/10.1103/PhysRevResearch.2.023151} {\bibfield  {journal} {\bibinfo  {journal} {Phys. Rev. Res.}\ }\textbf {\bibinfo {volume} {2}},\ \bibinfo {pages} {023151} (\bibinfo {year} {2020})},\ \Eprint {https://arxiv.org/abs/1912.10055} {arXiv:1912.10055 [gr-qc]} \BibitemShut {NoStop}%
\bibitem [{\citenamefont {Abbott}\ \emph {et~al.}(2017{\natexlab{b}})\citenamefont {Abbott} \emph {et~al.}}]{LIGOScientific:2017vwq}%
  \BibitemOpen
  \bibfield  {author} {\bibinfo {author} {\bibfnamefont {B.~P.}\ \bibnamefont {Abbott}} \emph {et~al.} (\bibinfo {collaboration} {LIGO Scientific, Virgo}),\ }\bibfield  {title} {\bibinfo {title} {{GW170817: Observation of Gravitational Waves from a Binary Neutron Star Inspiral}},\ }\href {https://doi.org/10.1103/PhysRevLett.119.161101} {\bibfield  {journal} {\bibinfo  {journal} {Phys. Rev. Lett.}\ }\textbf {\bibinfo {volume} {119}},\ \bibinfo {pages} {161101} (\bibinfo {year} {2017}{\natexlab{b}})},\ \Eprint {https://arxiv.org/abs/1710.05832} {arXiv:1710.05832 [gr-qc]} \BibitemShut {NoStop}%
\bibitem [{\citenamefont {Hanselman}\ \emph {et~al.}(2025)\citenamefont {Hanselman}, \citenamefont {Vijaykumar}, \citenamefont {Fishbach},\ and\ \citenamefont {Holz}}]{Hanselman:2024hqy}%
  \BibitemOpen
  \bibfield  {author} {\bibinfo {author} {\bibfnamefont {A.~G.}\ \bibnamefont {Hanselman}}, \bibinfo {author} {\bibfnamefont {A.}~\bibnamefont {Vijaykumar}}, \bibinfo {author} {\bibfnamefont {M.}~\bibnamefont {Fishbach}},\ and\ \bibinfo {author} {\bibfnamefont {D.~E.}\ \bibnamefont {Holz}},\ }\bibfield  {title} {\bibinfo {title} {{Gravitational-wave Dark Siren Cosmology Systematics from Galaxy Weighting}},\ }\href {https://doi.org/10.3847/1538-4357/ad9393} {\bibfield  {journal} {\bibinfo  {journal} {Astrophys. J.}\ }\textbf {\bibinfo {volume} {979}},\ \bibinfo {pages} {9} (\bibinfo {year} {2025})},\ \Eprint {https://arxiv.org/abs/2405.14818} {arXiv:2405.14818 [astro-ph.CO]} \BibitemShut {NoStop}%
\bibitem [{\citenamefont {Abbott}\ \emph {et~al.}(2023{\natexlab{b}})\citenamefont {Abbott} \emph {et~al.}}]{KAGRA:2021duu}%
  \BibitemOpen
  \bibfield  {author} {\bibinfo {author} {\bibfnamefont {R.}~\bibnamefont {Abbott}} \emph {et~al.} (\bibinfo {collaboration} {KAGRA, VIRGO, LIGO Scientific}),\ }\bibfield  {title} {\bibinfo {title} {{Population of Merging Compact Binaries Inferred Using Gravitational Waves through GWTC-3}},\ }\href {https://doi.org/10.1103/PhysRevX.13.011048} {\bibfield  {journal} {\bibinfo  {journal} {Phys. Rev. X}\ }\textbf {\bibinfo {volume} {13}},\ \bibinfo {pages} {011048} (\bibinfo {year} {2023}{\natexlab{b}})},\ \Eprint {https://arxiv.org/abs/2111.03634} {arXiv:2111.03634 [astro-ph.HE]} \BibitemShut {NoStop}%
\bibitem [{\citenamefont {Abbott}\ \emph {et~al.}(2020)\citenamefont {Abbott} \emph {et~al.}}]{LIGOScientific:2020iuh}%
  \BibitemOpen
  \bibfield  {author} {\bibinfo {author} {\bibfnamefont {R.}~\bibnamefont {Abbott}} \emph {et~al.} (\bibinfo {collaboration} {LIGO Scientific, Virgo}),\ }\bibfield  {title} {\bibinfo {title} {{GW190521: A Binary Black Hole Merger with a Total Mass of $150 M_{\odot}$}},\ }\href {https://doi.org/10.1103/PhysRevLett.125.101102} {\bibfield  {journal} {\bibinfo  {journal} {Phys. Rev. Lett.}\ }\textbf {\bibinfo {volume} {125}},\ \bibinfo {pages} {101102} (\bibinfo {year} {2020})},\ \Eprint {https://arxiv.org/abs/2009.01075} {arXiv:2009.01075 [gr-qc]} \BibitemShut {NoStop}%
\bibitem [{\citenamefont {Abbott}\ \emph {et~al.}(2023{\natexlab{c}})\citenamefont {Abbott} \emph {et~al.}}]{KAGRA:2021vkt}%
  \BibitemOpen
  \bibfield  {author} {\bibinfo {author} {\bibfnamefont {R.}~\bibnamefont {Abbott}} \emph {et~al.} (\bibinfo {collaboration} {KAGRA, VIRGO, LIGO Scientific}),\ }\bibfield  {title} {\bibinfo {title} {{GWTC-3: Compact Binary Coalescences Observed by LIGO and Virgo during the Second Part of the Third Observing Run}},\ }\href {https://doi.org/10.1103/PhysRevX.13.041039} {\bibfield  {journal} {\bibinfo  {journal} {Phys. Rev. X}\ }\textbf {\bibinfo {volume} {13}},\ \bibinfo {pages} {041039} (\bibinfo {year} {2023}{\natexlab{c}})},\ \Eprint {https://arxiv.org/abs/2111.03606} {arXiv:2111.03606 [gr-qc]} \BibitemShut {NoStop}%
\bibitem [{\citenamefont {Rinaldi}\ \emph {et~al.}(2024)\citenamefont {Rinaldi}, \citenamefont {Del~Pozzo}, \citenamefont {Mapelli}, \citenamefont {Lorenzo-Medina},\ and\ \citenamefont {Dent}}]{Rinaldi:2023bbd}%
  \BibitemOpen
  \bibfield  {author} {\bibinfo {author} {\bibfnamefont {S.}~\bibnamefont {Rinaldi}}, \bibinfo {author} {\bibfnamefont {W.}~\bibnamefont {Del~Pozzo}}, \bibinfo {author} {\bibfnamefont {M.}~\bibnamefont {Mapelli}}, \bibinfo {author} {\bibfnamefont {A.}~\bibnamefont {Lorenzo-Medina}},\ and\ \bibinfo {author} {\bibfnamefont {T.}~\bibnamefont {Dent}},\ }\bibfield  {title} {\bibinfo {title} {{Evidence of evolution of the black hole mass function with redshift}},\ }\href {https://doi.org/10.1051/0004-6361/202348161} {\bibfield  {journal} {\bibinfo  {journal} {Astron. Astrophys.}\ }\textbf {\bibinfo {volume} {684}},\ \bibinfo {pages} {A204} (\bibinfo {year} {2024})},\ \Eprint {https://arxiv.org/abs/2310.03074} {arXiv:2310.03074 [astro-ph.HE]} \BibitemShut {NoStop}%
\bibitem [{\citenamefont {Lalleman}\ \emph {et~al.}(2025)\citenamefont {Lalleman}, \citenamefont {Turbang}, \citenamefont {Callister},\ and\ \citenamefont {van Remortel}}]{Lalleman:2025xcs}%
  \BibitemOpen
  \bibfield  {author} {\bibinfo {author} {\bibfnamefont {M.}~\bibnamefont {Lalleman}}, \bibinfo {author} {\bibfnamefont {K.}~\bibnamefont {Turbang}}, \bibinfo {author} {\bibfnamefont {T.}~\bibnamefont {Callister}},\ and\ \bibinfo {author} {\bibfnamefont {N.}~\bibnamefont {van Remortel}},\ }\bibfield  {title} {\bibinfo {title} {{No evidence that the binary black hole mass distribution evolves with redshift}},\ }\href@noop {} {\  (\bibinfo {year} {2025})},\ \Eprint {https://arxiv.org/abs/2501.10295} {arXiv:2501.10295 [astro-ph.HE]} \BibitemShut {NoStop}%
\bibitem [{\citenamefont {Vink}\ \emph {et~al.}(2021)\citenamefont {Vink}, \citenamefont {Higgins}, \citenamefont {Sander},\ and\ \citenamefont {Sabhahit}}]{Vink:2020nak}%
  \BibitemOpen
  \bibfield  {author} {\bibinfo {author} {\bibfnamefont {J.~S.}\ \bibnamefont {Vink}}, \bibinfo {author} {\bibfnamefont {E.~R.}\ \bibnamefont {Higgins}}, \bibinfo {author} {\bibfnamefont {A.~A.~C.}\ \bibnamefont {Sander}},\ and\ \bibinfo {author} {\bibfnamefont {G.~N.}\ \bibnamefont {Sabhahit}},\ }\bibfield  {title} {\bibinfo {title} {{Maximum black hole mass across cosmic time}},\ }\href {https://doi.org/10.1093/mnras/stab842} {\bibfield  {journal} {\bibinfo  {journal} {Mon. Not. Roy. Astron. Soc.}\ }\textbf {\bibinfo {volume} {504}},\ \bibinfo {pages} {146} (\bibinfo {year} {2021})},\ \Eprint {https://arxiv.org/abs/2010.11730} {arXiv:2010.11730 [astro-ph.HE]} \BibitemShut {NoStop}%
\bibitem [{\citenamefont {Weatherford}\ \emph {et~al.}(2021)\citenamefont {Weatherford}, \citenamefont {Fragione}, \citenamefont {Kremer}, \citenamefont {Chatterjee}, \citenamefont {Ye}, \citenamefont {Rodriguez},\ and\ \citenamefont {Rasio}}]{Weatherford:2021zdf}%
  \BibitemOpen
  \bibfield  {author} {\bibinfo {author} {\bibfnamefont {N.~C.}\ \bibnamefont {Weatherford}}, \bibinfo {author} {\bibfnamefont {G.}~\bibnamefont {Fragione}}, \bibinfo {author} {\bibfnamefont {K.}~\bibnamefont {Kremer}}, \bibinfo {author} {\bibfnamefont {S.}~\bibnamefont {Chatterjee}}, \bibinfo {author} {\bibfnamefont {C.~S.}\ \bibnamefont {Ye}}, \bibinfo {author} {\bibfnamefont {C.~L.}\ \bibnamefont {Rodriguez}},\ and\ \bibinfo {author} {\bibfnamefont {F.~A.}\ \bibnamefont {Rasio}},\ }\bibfield  {title} {\bibinfo {title} {{Black Hole Mergers from Star Clusters with Top-Heavy Initial Mass Functions}},\ }\href {https://doi.org/10.3847/2041-8213/abd79c} {\bibfield  {journal} {\bibinfo  {journal} {Astrophys. J. Lett.}\ }\textbf {\bibinfo {volume} {907}},\ \bibinfo {pages} {L25} (\bibinfo {year} {2021})},\ \Eprint {https://arxiv.org/abs/2101.02217} {arXiv:2101.02217 [astro-ph.GA]} \BibitemShut {NoStop}%
\bibitem [{\citenamefont {Zevin}\ \emph {et~al.}(2021)\citenamefont {Zevin}, \citenamefont {Bavera}, \citenamefont {Berry}, \citenamefont {Kalogera}, \citenamefont {Fragos}, \citenamefont {Marchant}, \citenamefont {Rodriguez}, \citenamefont {Antonini}, \citenamefont {Holz},\ and\ \citenamefont {Pankow}}]{Zevin:2020gbd}%
  \BibitemOpen
  \bibfield  {author} {\bibinfo {author} {\bibfnamefont {M.}~\bibnamefont {Zevin}}, \bibinfo {author} {\bibfnamefont {S.~S.}\ \bibnamefont {Bavera}}, \bibinfo {author} {\bibfnamefont {C.~P.~L.}\ \bibnamefont {Berry}}, \bibinfo {author} {\bibfnamefont {V.}~\bibnamefont {Kalogera}}, \bibinfo {author} {\bibfnamefont {T.}~\bibnamefont {Fragos}}, \bibinfo {author} {\bibfnamefont {P.}~\bibnamefont {Marchant}}, \bibinfo {author} {\bibfnamefont {C.~L.}\ \bibnamefont {Rodriguez}}, \bibinfo {author} {\bibfnamefont {F.}~\bibnamefont {Antonini}}, \bibinfo {author} {\bibfnamefont {D.~E.}\ \bibnamefont {Holz}},\ and\ \bibinfo {author} {\bibfnamefont {C.}~\bibnamefont {Pankow}},\ }\bibfield  {title} {\bibinfo {title} {{One Channel to Rule Them All? Constraining the Origins of Binary Black Holes Using Multiple Formation Pathways}},\ }\href {https://doi.org/10.3847/1538-4357/abe40e} {\bibfield  {journal} {\bibinfo  {journal} {Astrophys. J.}\ }\textbf {\bibinfo {volume} {910}},\ \bibinfo {pages} {152} (\bibinfo {year} {2021})},\ \Eprint {https://arxiv.org/abs/2011.10057} {arXiv:2011.10057 [astro-ph.HE]} \BibitemShut {NoStop}%
\bibitem [{\citenamefont {Bavera}\ \emph {et~al.}(2022)\citenamefont {Bavera}, \citenamefont {Fishbach}, \citenamefont {Zevin}, \citenamefont {Zapartas},\ and\ \citenamefont {Fragos}}]{Bavera:2022mef}%
  \BibitemOpen
  \bibfield  {author} {\bibinfo {author} {\bibfnamefont {S.~S.}\ \bibnamefont {Bavera}}, \bibinfo {author} {\bibfnamefont {M.}~\bibnamefont {Fishbach}}, \bibinfo {author} {\bibfnamefont {M.}~\bibnamefont {Zevin}}, \bibinfo {author} {\bibfnamefont {E.}~\bibnamefont {Zapartas}},\ and\ \bibinfo {author} {\bibfnamefont {T.}~\bibnamefont {Fragos}},\ }\bibfield  {title} {\bibinfo {title} {{The {\ensuremath{\chi}}eff {\ensuremath{-}} z correlation of field binary black hole mergers and how 3G gravitational-wave detectors can constrain it}},\ }\href {https://doi.org/10.1051/0004-6361/202243724} {\bibfield  {journal} {\bibinfo  {journal} {Astron. Astrophys.}\ }\textbf {\bibinfo {volume} {665}},\ \bibinfo {pages} {A59} (\bibinfo {year} {2022})},\ \Eprint {https://arxiv.org/abs/2204.02619} {arXiv:2204.02619 [astro-ph.HE]} \BibitemShut {NoStop}%
\bibitem [{\citenamefont {Biscoveanu}\ \emph {et~al.}(2022)\citenamefont {Biscoveanu}, \citenamefont {Callister}, \citenamefont {Haster}, \citenamefont {Ng}, \citenamefont {Vitale},\ and\ \citenamefont {Farr}}]{Biscoveanu:2022qac}%
  \BibitemOpen
  \bibfield  {author} {\bibinfo {author} {\bibfnamefont {S.}~\bibnamefont {Biscoveanu}}, \bibinfo {author} {\bibfnamefont {T.~A.}\ \bibnamefont {Callister}}, \bibinfo {author} {\bibfnamefont {C.-J.}\ \bibnamefont {Haster}}, \bibinfo {author} {\bibfnamefont {K.~K.~Y.}\ \bibnamefont {Ng}}, \bibinfo {author} {\bibfnamefont {S.}~\bibnamefont {Vitale}},\ and\ \bibinfo {author} {\bibfnamefont {W.~M.}\ \bibnamefont {Farr}},\ }\bibfield  {title} {\bibinfo {title} {{The Binary Black Hole Spin Distribution Likely Broadens with Redshift}},\ }\href {https://doi.org/10.3847/2041-8213/ac71a8} {\bibfield  {journal} {\bibinfo  {journal} {Astrophys. J. Lett.}\ }\textbf {\bibinfo {volume} {932}},\ \bibinfo {pages} {L19} (\bibinfo {year} {2022})},\ \Eprint {https://arxiv.org/abs/2204.01578} {arXiv:2204.01578 [astro-ph.HE]} \BibitemShut {NoStop}%
\bibitem [{\citenamefont {Ye}\ and\ \citenamefont {Fishbach}(2024)}]{Ye:2024ypm}%
  \BibitemOpen
  \bibfield  {author} {\bibinfo {author} {\bibfnamefont {C.~S.}\ \bibnamefont {Ye}}\ and\ \bibinfo {author} {\bibfnamefont {M.}~\bibnamefont {Fishbach}},\ }\bibfield  {title} {\bibinfo {title} {{The Redshift Evolution of the Binary Black Hole Mass Distribution from Dense Star Clusters}},\ }\href {https://doi.org/10.3847/1538-4357/ad3ba8} {\bibfield  {journal} {\bibinfo  {journal} {Astrophys. J.}\ }\textbf {\bibinfo {volume} {967}},\ \bibinfo {pages} {62} (\bibinfo {year} {2024})},\ \Eprint {https://arxiv.org/abs/2402.12444} {arXiv:2402.12444 [astro-ph.HE]} \BibitemShut {NoStop}%
\bibitem [{\citenamefont {Gair}\ and\ \citenamefont {Moore}(2015)}]{Gair:2015nga}%
  \BibitemOpen
  \bibfield  {author} {\bibinfo {author} {\bibfnamefont {J.~R.}\ \bibnamefont {Gair}}\ and\ \bibinfo {author} {\bibfnamefont {C.~J.}\ \bibnamefont {Moore}},\ }\bibfield  {title} {\bibinfo {title} {{Quantifying and mitigating bias in inference on gravitational wave source populations}},\ }\href {https://doi.org/10.1103/PhysRevD.91.124062} {\bibfield  {journal} {\bibinfo  {journal} {Phys. Rev. D}\ }\textbf {\bibinfo {volume} {91}},\ \bibinfo {pages} {124062} (\bibinfo {year} {2015})},\ \Eprint {https://arxiv.org/abs/1504.02767} {arXiv:1504.02767 [gr-qc]} \BibitemShut {NoStop}%
\bibitem [{\citenamefont {Isi}\ \emph {et~al.}(2022)\citenamefont {Isi}, \citenamefont {Farr},\ and\ \citenamefont {Chatziioannou}}]{Isi:2022cii}%
  \BibitemOpen
  \bibfield  {author} {\bibinfo {author} {\bibfnamefont {M.}~\bibnamefont {Isi}}, \bibinfo {author} {\bibfnamefont {W.~M.}\ \bibnamefont {Farr}},\ and\ \bibinfo {author} {\bibfnamefont {K.}~\bibnamefont {Chatziioannou}},\ }\bibfield  {title} {\bibinfo {title} {{Comparing Bayes factors and hierarchical inference for testing general relativity with gravitational waves}},\ }\href {https://doi.org/10.1103/PhysRevD.106.024048} {\bibfield  {journal} {\bibinfo  {journal} {Phys. Rev. D}\ }\textbf {\bibinfo {volume} {106}},\ \bibinfo {pages} {024048} (\bibinfo {year} {2022})},\ \Eprint {https://arxiv.org/abs/2204.10742} {arXiv:2204.10742 [gr-qc]} \BibitemShut {NoStop}%
\bibitem [{\citenamefont {Dax}\ \emph {et~al.}(2022)\citenamefont {Dax}, \citenamefont {Green}, \citenamefont {Gair}, \citenamefont {P\"urrer}, \citenamefont {Wildberger}, \citenamefont {Macke}, \citenamefont {Buonanno},\ and\ \citenamefont {Sch\"olkopf}}]{Dax:2022pxd}%
  \BibitemOpen
  \bibfield  {author} {\bibinfo {author} {\bibfnamefont {M.}~\bibnamefont {Dax}}, \bibinfo {author} {\bibfnamefont {S.~R.}\ \bibnamefont {Green}}, \bibinfo {author} {\bibfnamefont {J.}~\bibnamefont {Gair}}, \bibinfo {author} {\bibfnamefont {M.}~\bibnamefont {P\"urrer}}, \bibinfo {author} {\bibfnamefont {J.}~\bibnamefont {Wildberger}}, \bibinfo {author} {\bibfnamefont {J.~H.}\ \bibnamefont {Macke}}, \bibinfo {author} {\bibfnamefont {A.}~\bibnamefont {Buonanno}},\ and\ \bibinfo {author} {\bibfnamefont {B.}~\bibnamefont {Sch\"olkopf}},\ }\bibfield  {title} {\bibinfo {title} {{Neural Importance Sampling for Rapid and Reliable Gravitational-Wave Inference}},\ }\href {https://doi.org/https://doi.org/10.1103/PhysRevLett.130.171403} {\bibfield  {journal} {\bibinfo  {journal} {Physical Review Letters}\ }\textbf {\bibinfo {volume} {130}},\ \bibinfo {pages} {171403} (\bibinfo {year} {2022})},\ \Eprint {https://arxiv.org/abs/2210.05686} {arXiv:2210.05686 [gr-qc]} \BibitemShut {NoStop}%
\end{thebibliography}%

\clearpage

\section*{Supplementary materials}
\appendix

\section{\mice}
\label{sec:micecat}
We assign \ac{GW} sources to galaxies taken from \mice, the Grand Challenge. 
It is a mock galaxy catalog that covers one octant of the sky up to a redshift of 1.4.
While the full catalog contains $\sim 205 \rm M$ galaxies,
we only consider galaxies with luminosity $L > 10^{10} L_{\odot}$, where $L_{\odot}$ 
is the solar luminosity, as possible hosts reducing the catalog size to $\sim 83 \rm M$ galaxies.
The $H_0$ calculation is done using this reduced catalog for consistency.
The fiducial cosmological model in \mice is a flat $\Lambda$CDM model with parameters 
$H_0 = 70 \rm \, km \, s^{-1} \, Mpc^{-1}$ and $\Omega_m = 0.25$. 
We choose the same model for simulating \ac{GW} sources.

\section{GW population}
\label{sec:HMP_population}
With close to 100 observations, we have an understanding of how the binary merger population is distributed.
We simulate two populations of \ac{BBH} mergers with 100k binaries in each. 
All parameters are the same across both the populations except the masses and spins.
The host galaxy for each merger is assigned by randomly drawing galaxies from \mice.
This specifies the redshift distribution and sky positions of the two populations.
Since the \mice mock catalog covers only one-octant of the sky the simulated \ac{GW} sources 
are also restricted to this single octant in the sky. 
We will always consider \ac{GW} networks with at least 3 non-colocated detectors and, therefore,
we expect each octant in the sky to give similar results since the network sensitivity for tensor polarizations
does not vary much across the sky. 
A binary orbit is assumed to be randomly oriented, and the polarization angle 
and orbital phase are also chosen randomly. The time of coalescence is fixed to some fiducial value.

In the first population, the masses and spins are chosen to follow \mass and \spin distributions of GWTC-3, respectively.
The motivation for the second population, denoted ``high-mass spin-precessing'', is to study the impact of an yet unobserved subpopulation consisting 
of binaries with large masses, large mass ratios, and large precessing spins. 
The impact of such a subpopulation on the astrophysical and cosmological inference using \acp{GW} is of interest 
to the waveform modelling community because it lies in a region of the parameter space that is relatively poorly
modelled. 
On the astrophysical side, such a population is of interest because the hierarchical merger scenario can produce them. 
We sample uniformly in the total source-frame mass and the inverse mass ratio in the interval 
$[10, 200] \rm M_{\odot}$ and $[1, 30]$, respectively, and place a constraint that the smaller mass is always greater 
than $5 \rm M_{\odot}$. 
This results in a non-uniform distribution of the two masses with a preference for larger masses. 
The spin magnitudes are sampled from a uniform distribution spanning its domain and the spin angles are 
isotropically distributed. 
Subsequently, this sample is subsampled such that the $\chi_p$ distribution is uniform.

\begin{figure*}
	\includegraphics[width=2\columnwidth]{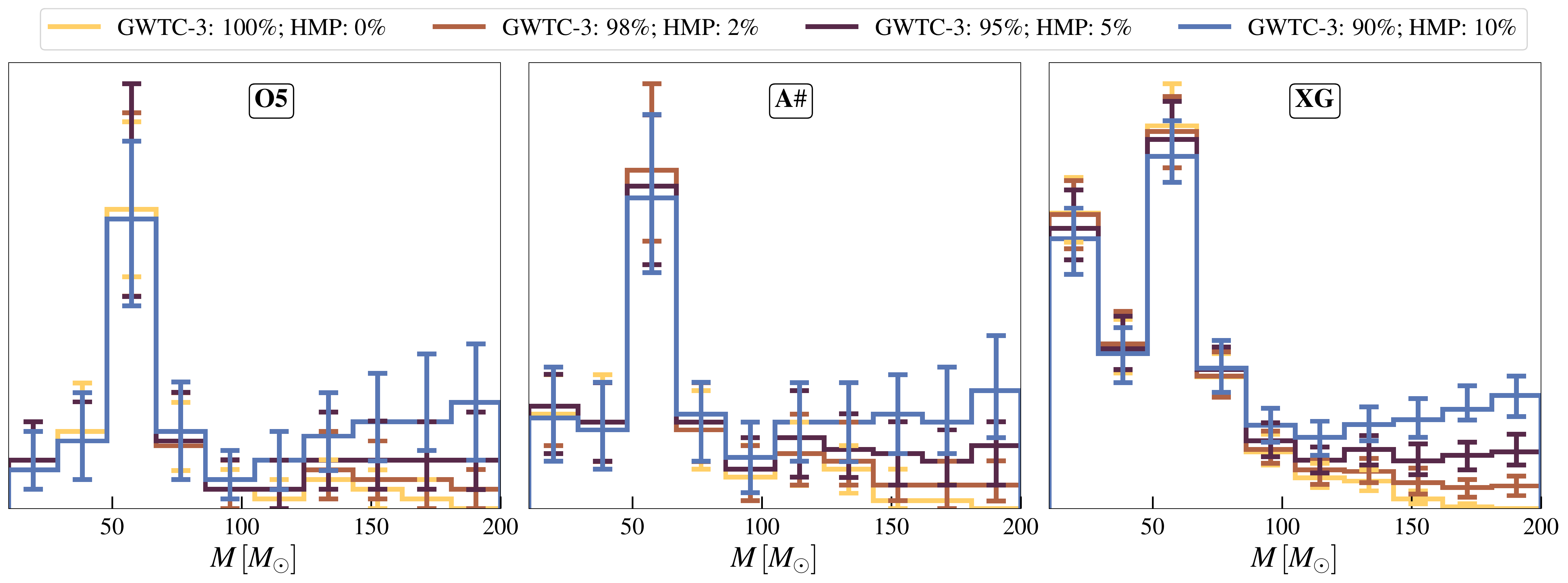}
	\includegraphics[width=2\columnwidth]{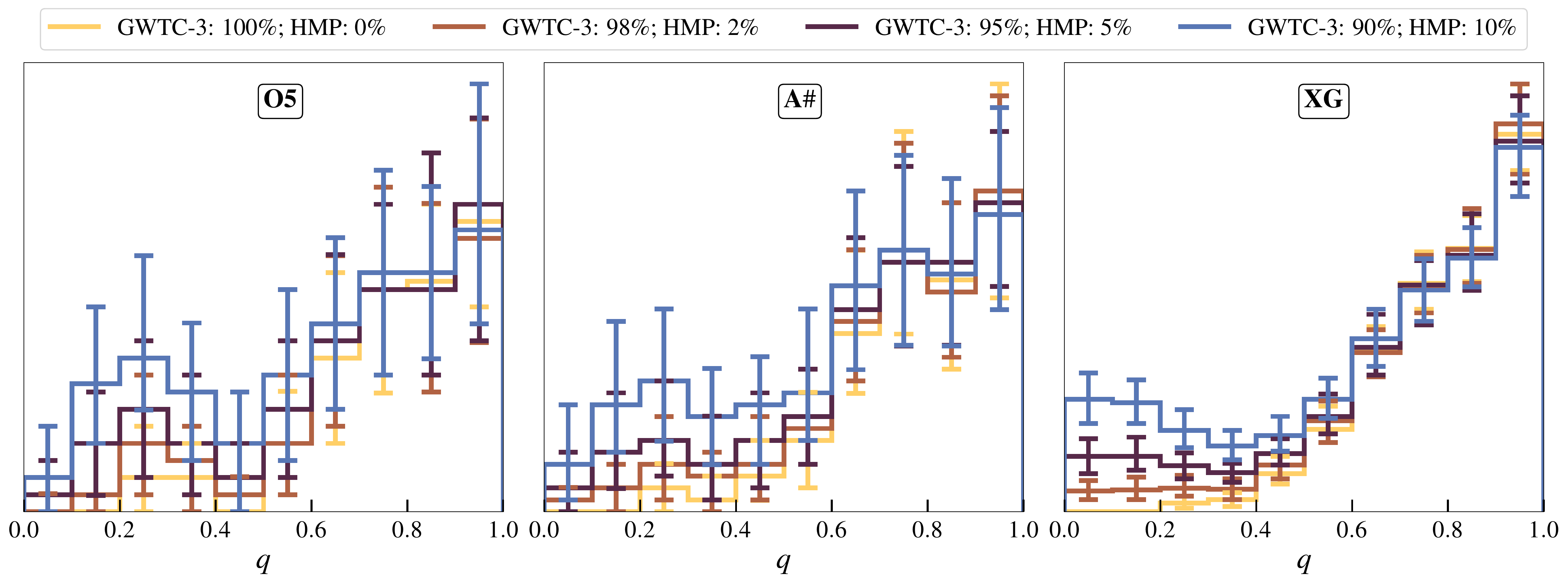}
	\includegraphics[width=2\columnwidth]{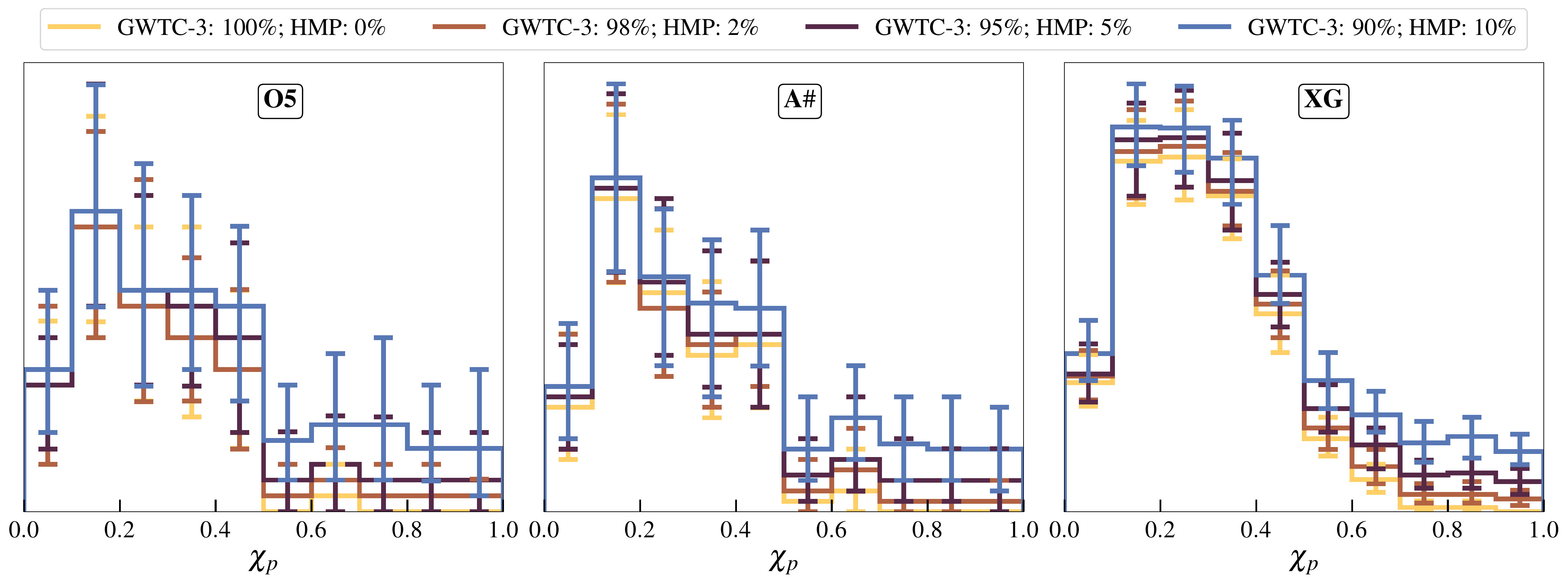}
	\caption{Mixture population crossing SNR threshold}
	\label{fig:mixture_pop}
\end{figure*}

The final simulated populations are linear combinations of these two underlying populations with varying relative weights. The distributions of the total mass, mass ratio, and spin-precession parameter are shown in \cref{fig:mixture_pop}.

\section{$H_0$ inference}
Consider a set of $N_{obs}$ events with single event parameters $\{\bm{\vartheta}_i\}_{i=1}^{N_{obs}}$ and 
data $\mathcal{D} = \{\mathcal{D}_{GW}^i, \mathcal{D}_{EM}^i\}_{i=1}^{N_{obs}}$. 
The \ac{EM} data, consisting of the mock galaxy catalog \mice, contains the redshift, position in the sky, and luminosity
of all galaxies in the catalog.
While one expects the hosting probability to depend on galaxy properties like its luminosity or 
star formation rate, \ac{GW} observations do not provide a clear answer yet. 
For quasi-circular, spin-precessing systems considered here, the \ac{GW} data depends on the usual 15 parameters:
component masses, component spins, location, orientation, and phase.
In general, one expects the mass and spin distributions to evolve with redshift and, therefore,
encode information about the true cosmology.
However, current \ac{GW} observations are uncertain of the form of such a dependence.
Even in the absence of a redshift evolution of the mass distribution, the deviation in the observed 
detector-frame mass distribution from the source-frame mass distribution encodes the redshift, 
a method that exploits this is known as \textit{spectral siren}~\cite{Chernoff:1993th,Taylor:2012db}.
Nevertheless, since we are interested in investigating $H_0$ systematics while using the galaxy catalog method, 
it is beneficial to marginalize over all the intrinsic parameters to isolate the effect of the galaxy catalog.
We are, in effect, using less information than what is available to us and this results in the $H_0$ posteriors 
being marginally broader than what they could be.

The contribution of a single event to the likelihood of $H_0$, after marginalisation of the joint likelihood over the number of detections using a log-uniform prior, is given by 
\begin{equation}
\begin{split}
	\mathcal{L}(\mathcal{D}_{GW}^i, &\mathcal{D}_{EM}^i|H_0) = \frac{1}{\beta(H_0)} \times \\
	& \int dz d\varTheta \mathcal{L}_{GW}(\mathcal{D}_{GW}^i|d_L(z,H_0),\varTheta) p(z,\varTheta),
	\label{eq:ind_lk}
\end{split}
\end{equation}
where $d_L$ and $\varTheta$ are the luminosity distance and sky position, respectively, $\mathcal{L}_{GW}(.)$ is the \ac{GW} likelihood, marginalized over all other parameters, and $p(z,\varTheta)$ is the joint probability distribution for the redshift and sky position of \ac{GW} sources 
obtained from the galaxy catalog. 
The denominator, $\beta(H_0)$, is a normalization constant given by,
\begin{equation}
	\beta(H_0) = \int d\mathcal{D}_{GW}^i dz d\varTheta \mathcal{L}_{GW}(\mathcal{D}_{GW}^i|d_L(z,H_0),\varTheta) p(z,\varTheta),
\end{equation}
where the $\mathcal{D}_{GW}^i$ integration is over all data that is above the detection criteria. This term corrects for selection effects. 
Here, we have assumed that the detection criteria is based on the \ac{GW} data alone and the galaxy catalog is complete.

We simplify \cref{eq:ind_lk} by assuming that the sky location and luminosity distance separate in the \ac{GW} likelihood and the 
redshift prior. 
The \ac{GW} likelihood then becomes  
$\mathcal{L}_{GW}(\mathcal{D}_{GW}^i|d_L(z,H_0),\varTheta) = p(\hat{d}_L^i|d_L(z,H_0)) p(\hat{\varTheta}^i|\varTheta)$
where $\hat{d}_L^i$ and $\hat{\varTheta}^i$ are the observed luminosity distance and sky location 
of a \ac{GW} event. 
We simulate an observed value by drawing a point-estimate from a Normal distribution centred at the 
injected value and covariance given by the inverse of the Fisher matrix.
Furthermore, we regard the \ac{EM} sky positions to be precise and set these to $\delta$-functions.
Under these assumptions, \cref{eq:ind_lk} becomes 
\begin{equation}
	\mathcal{L}(\mathcal{D}^i|H_0) = \frac{1}{\beta(H_0)} \sum_{j=1}^{N_{gal}}
	\int dz_j \mathcal{N}(\hat{d}_L^i|d_L(z_j,H_0))  w_j^i p(z_j)
	\label{eq:h0_lk}
\end{equation}
where $w_j^i = \mathcal{N}(\hat{\varTheta}^i|\varTheta_0^j,\Sigma^i)$ with $\varTheta_0^j$ the position 
of the j-th galalxy in the galaxy catalog and $N_{gal}$ the total number of galaxies in the localization
volume. 
$p(z_j)$ is the posterior distribution on the redshift. 
We assume the redshifts to be known precisely and model it as a $\delta$-function.
This is a reasonable assumption if spectroscopic measurements of the redshifts of the galaxies 
are available.
On the contrary, photometric measurements have larger measurement errors which would need to be 
taken into account. 

\section{Population diagnostic}
\label{sec:population_diagnostic}
The posterior distribution on the hyperparameters $\mu, \sigma$ is given by
\begin{equation}
	p(\mu,\sigma|\{\mathcal{D}\}) = p(\mu,\sigma) \prod_{i=1}^{N_{obs}} \int dH_0^i \mathcal{L}(\mathcal{D}^i|H_0^i) \mathcal{N}(H_0^i|\mu,\sigma)
\end{equation}
where the $H_0$ likelihood is defined in \cref{eq:h0_lk}. 
We define the \ac{PPHD} on \ac{$H_0$} as 
\begin{equation}
	p(H_0|\{\mathcal{D}\}) = \int d\mu d\sigma \mathcal{N}(H_0|\mu,\sigma) p(\mu,\sigma|\{\mathcal{D}\}).
\end{equation}

\section{\ac{PPHD}}
\begin{figure}
	\includegraphics[width=\columnwidth]{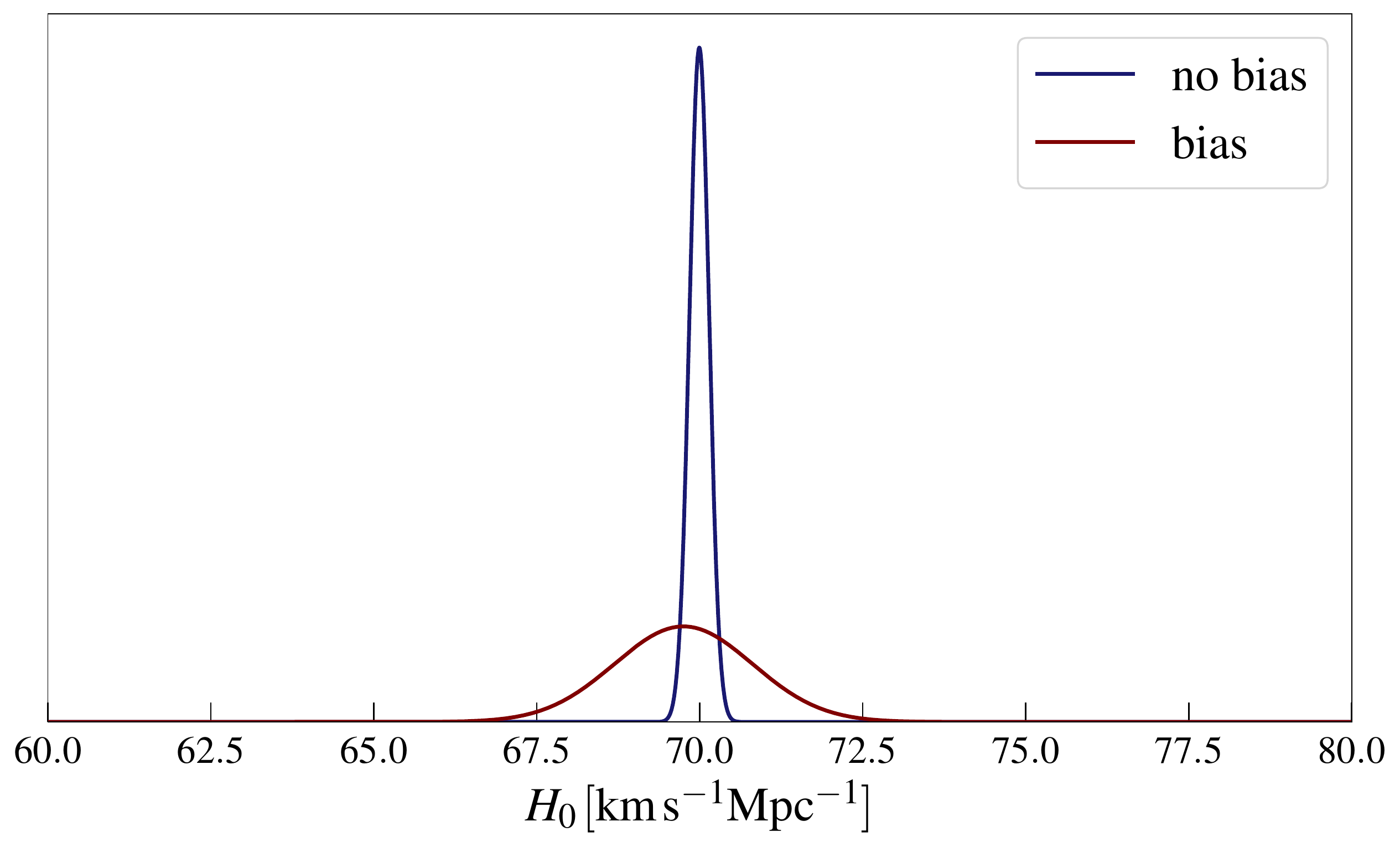}
	\caption{The posterior predictive distributions (PPD) of the population parameter $H_0$ for a GWTC-3 population
	in the \XG network. 
	The PPD, in the absence of systematic biases, tends toward a $\delta-$function, as expected. 
	On the other hand, a non-zero $\sigma$ causes the PPD to have a finite width, as in the presence of systematic biases.}
	\label{fig:ppd}
\end{figure}

In \cref{fig:ppd}, we show the \ac{PPHD} for the GWTC-3 population in \XG network. 
The distribution tends towards a $\delta$-function in the absence of systematic biases but asymptotes to a finitely wide 
distribution otherwise.

\begin{figure}
	\includegraphics[width=\columnwidth]{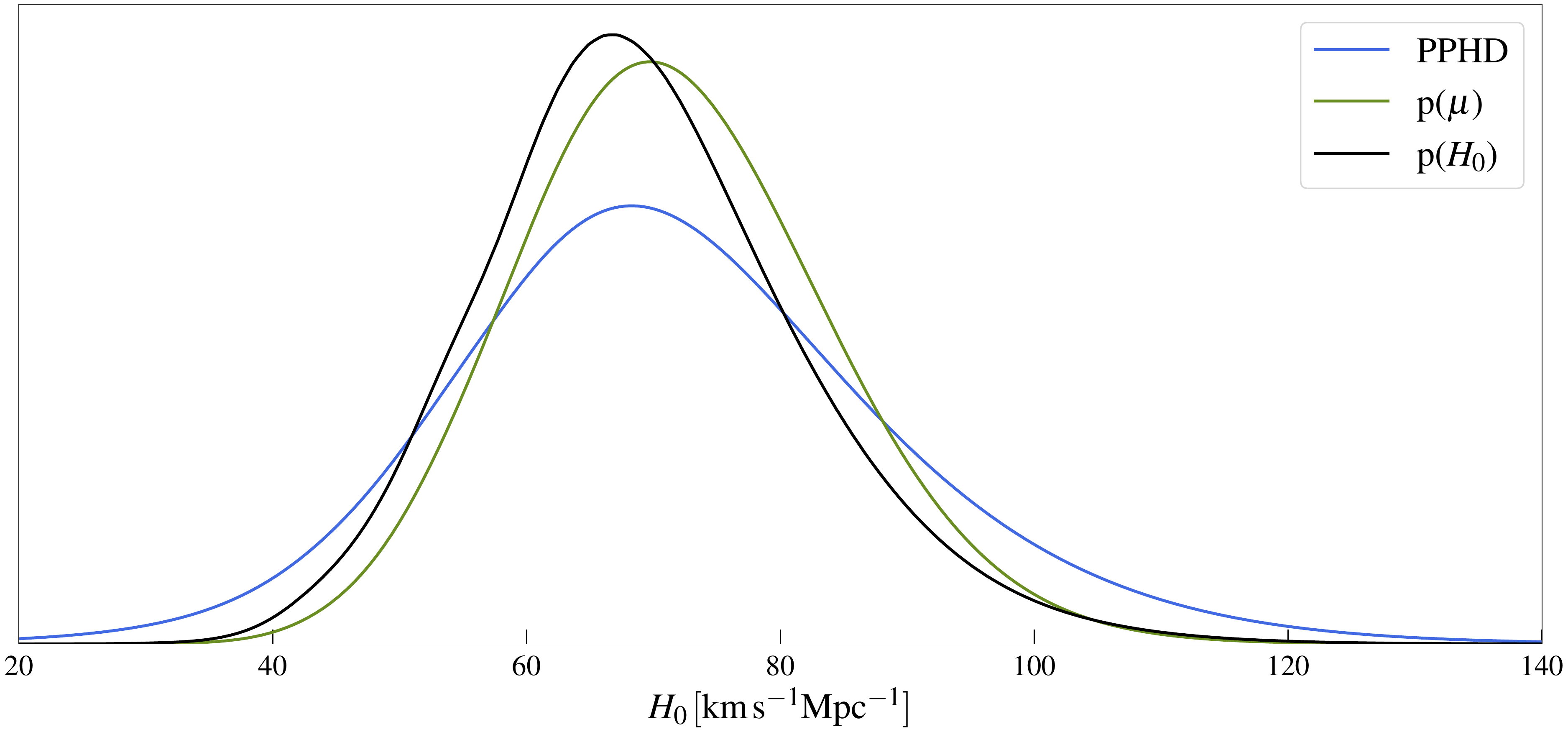}
	\caption{The posterior predictive distribution, the marginalized $\mu$ distribution, and the joint posterior 
	distribution on $H_0$ obtained from the events used in the GWTC-3 cosmology paper~\cite{LIGOScientific:2021aug}. 
	We do not see any indication of systematic biases. 
	Note that even though the $\mu$ posterior is very similar to the joint $H_0$ posterior but we are not yet in 
	the large number of observations limit where the two will coincide.}
	\label{fig:gwtc3_ppd}
\end{figure}

In \cref{fig:gwtc3_ppd}, we show the joint $H_0$ posterior (which is the $\mu$ posterior conditioned on $\sigma=0$), 
the marginalised $\mu$ posterior and the \ac{PPHD} for the 
real events analzed in the GWTC-3 cosmology paper \cite{LIGOScientific:2021aug}.

\end{document}